\documentclass[12pt]{article}
\usepackage{graphicx}
\graphicspath{{SynthIPD_figures/}}
\usepackage{enumerate}
\usepackage[numbers]{natbib}
\usepackage{amsthm,amsmath,amsfonts,amssymb,bbm,bm,enumitem,tikz,verbatim,multirow,multicol,xcolor,booktabs,array,tabularx}
\DeclareMathOperator*{\argmin}{argmin}

\usetikzlibrary{arrows.meta, positioning, calc, shadows.blur}

\definecolor{labelTeal}{HTML}{2F5D62}
\definecolor{labelNavy}{HTML}{2A3F64}
\definecolor{labelGold}{HTML}{B68B3A}
\definecolor{arrowNavy}{HTML}{2A3F64}
\definecolor{stepOrange}{HTML}{D9895E}
\definecolor{stepBlue}{HTML}{3E7CB1}
\definecolor{headBg}{HTML}{E3E6EE}
\definecolor{rowAlt}{HTML}{F6F6F9}
\definecolor{sectionLbl}{HTML}{444955}

\newcommand{\ipdtable}{%
  \renewcommand{\arraystretch}{1.35}%
  \setlength{\tabcolsep}{8pt}%
  \small
  \begin{tabular}{|c|c|c|c|c|}
    \hline 
    \textbf{Patient ID} & \textbf{Survival time} & \textbf{Status}
    & \textbf{Treatment} & \textbf{Covariate 1 (Gender)} \\ \hline
    1   & 1.5 mo & Event & FOLFOX & Female \\ \hline
    2   & 0.8 mo & Event & IROX   & Male \\ \hline
    $\cdots$ & $\cdots$ & $\cdots$ & $\cdots$ & $\cdots$ \\ \hline
  \end{tabular}%
}
 
\newcommand{\subgrouptable}{%
  \renewcommand{\arraystretch}{1.2}%
  \setlength{\tabcolsep}{5pt}%
  \small
  \begin{tabular}{@{}l cc cc c@{}}
    \toprule
    & \multicolumn{2}{c}{\textbf{FOLFOX} ($n = 421$)}
    & \multicolumn{2}{c}{\textbf{IROX} ($n = 383$)}
    & \textbf{HR (95\% CI)} \\
    \cmidrule(lr){2-3}\cmidrule(lr){4-5}
    & Events/Total & Median PFS (95\% CI)
    & Events/Total & Median PFS (95\% CI)
    & IROX vs FOLFOX \\
    \midrule
    \textbf{Gender} & & & & & \\
    \quad Male   & 241/247 & 8.2 (7.8--9.4)  & 225/230 & 6.5 (5.8--7.8) & 1.26 (1.05--1.51) \\
    \quad Female & 170/174 & 9.7 (8.5--11.0) & 149/153 & 6.9 (5.7--8.6) & 1.29 (1.04--1.61) \\
    \bottomrule
  \end{tabular}%
}
   
\usepackage{subfigure}
\usepackage{url} % not crucial - just used below for the URL 
\usepackage[normalem]{ulem}

\usepackage{algorithm}
\usepackage{algpseudocode}
\usepackage{algorithmicx}
\tikzstyle{process} = [rectangle, 
text centered, 
minimum width=2cm, 
text width=5cm,
draw = black]

\usepackage{tikz}
\usetikzlibrary{arrows.meta, positioning, shapes, fit}

\newcommand{\blind}{1}
\theoremstyle{plain}
\newtheorem{definition}{Definition}
\newtheorem{prop}{Proposition}
\newtheorem{example}{Example}

\newtheorem*{remark}{Remark}

\newtheorem{condition}{Condition}
\newtheorem{theorem}{Theorem}

\newcommand{\tHR}{\text{HR}}
\newcommand{\vX}{\bm{X}}

\begin{document}

\def\spacingset#1{\renewcommand{\baselinestretch}%
{#1}\small\normalsize} \spacingset{1}

%%%%%%%%%%%%%%%%%%%%%%%%%%%%%%%%%%%%%%%%%%%%%%%%%%%%%%%%%%%%%%%%%%%%%%%%%%%%%%

\if1\blind
{
  \title{\bf SynthIPD: training-free synthetic individual patient data generation}
  \author{Zixuan Zhao\thanks{Co-first author.}, Zexin Ren\footnotemark[1], Guannan Zhai and Feifang Hu\thanks{Corresponding author. (Email: feifang@gwu.edu)}
    % The authors gratefully acknowledge \textit{please remember to list all relevant funding sources in the unblinded version}}\hspace{.2cm}
    \\
    Department of Statistics, The George Washington University\\
    Will Ma, En Xie, HopeAI, Inc. \\
    Qian Shi, Mayo Clinic
    }
  \maketitle
} \fi

\if0\blind
{
  \bigskip
  \bigskip
  \bigskip
  \begin{center}
    {\LARGE\bf SynthIPD: training-free synthetic individual patient data generation}
\end{center}
  \medskip
} \fi

\bigskip
\begin{abstract}
Individual patient data (IPD) are essential for statistical analysis in clinical research, yet access is often limited due to privacy concerns, high data-sharing costs, and proprietary restrictions. Conventional approaches to synthetic data generation, such as generative adversarial networks (GANs), require a large piece of IPD as training set. We propose a training-free, three-step method to create synthetic IPD for survival data, requiring no IPD for training. In summary, it digitizes the reported Kaplan-Meier (KM) plots and generates covariates that match the reconstructed data. Compared with existing IPD reconstruction methods, our approach is the first to exploit Scalable Vector Graphics (SVG) for high-accuracy digitization, and the first to provide covariate information. We demonstrate the method’s potential through two detailed case studies and complementary simulation studies. 

% The proposed methodology offers important implications to evidence-based medical decision-making.
\end{abstract}

\noindent%
{\it Keywords:} Individual patient data, Kaplan-Meier curve, Synthetic data generation with covariates,   survival analysis, Vector Graphics
\vfill

\newpage
\spacingset{1.9} % DON'T change the spacing!

\section{Introduction}\label{sec:intro}
Individual patient data (IPD) sets are raw data collected from each patient and are usually essential for statistical analyses in clinical trials. Recent U.S. Food and Drug Administration guidelines on the use of Artificial Intelligence (AI) \cite{fda2025guidance} states that ``\textit{Continuous advancements in AI (to produce information or data) hold the potential to accelerate the development of safe and effective drugs and enhance patient care.}'' One out of many such advancements is synthetic IPD generation by AI, including machine learning models, which unlocks the potential of more efficient trial designs and supports regulatory decision-making. The guideline \cite{fda2025guidance} also stresses ``\textit{The variability in the quality, size, and representatives of datasets for training AI models may introduce bias and raise questions \ldots another challenge is the  model's performance to change over time when new data are introduced.}''

We offer a new perspective to the synthetic data problem for survival data. This work introduces SynthIPD, an approach to generate synthetic IPD \textbf{without relying on training datasets}. First, the coordinates of each event are digitized with high accuracy from published Kaplan-Meier (KM) plots and transformed into survival data. This step is done utilizing Scalable Vector Graphics (SVG), a picture format for describing two-dimensional graphics. The structure of SVG images allows them to be scaled to any resolution without loss of clarity, making them ideal for reconstruction by computer programming. Second, covariate data are generated via a constrained optimization approach. Finally, the synthetic survival and covariate data are combined, producing a final synthetic IPD.  As said, the proposed method is free of any statistical modeling assumptions because it is free from any training. However, it still requires some practical assumptions on its input, which will be discussed in Section \ref{subsec:synthipd}.

\subsection{Key applications}
SynthIPD enables numerous applications in biomedical studies. Some, out of many, are listed here: (i) Gaining new statistical insights: SynthIPD closely follows the distribution of the true IPD and can offer new statistical insights that are not originally reported in the publication \cite{thorlund2020synthetic}.  (ii) Establishing control benchmarks: By pooling synthetic IPD from the control arms of historical trials, investigators can establish an accurate median survival estimation for the control arm and further guide subsequent sample size calculation \cite{zhang2010calculating,lee2001uniform,zhu2016sample}. Such practice is crucial for the planning of novel efficient clinical trial design with covariates, e.g., covariate-adaptive randomization \cite{lin2015pursuit,hu2014adaptive,pocock1975sequential,huhu2012,rosenberger2012adaptive}. (iii) Constructing synthetic control arms: SynthIPD offers a potential way to reduce sample size by replacing or augmenting concurrent control arms. Synthetic clinical trial data, as pointed out by \cite{fda2023considerations}, have advantages over other sources, including real world data. (iv) Improving aggregated analyses: Pooled analyses using IPD can ``\textit{improve the quality of meta-analyses\ldots possibly correct for any deficiencies identified}'' \cite[Section H]{guidelinefdameta}, and is always ``\textit{preferred over aggregate data, especially for subgroups}'' \cite[Section 3.6]{ema2024indirect}. SynthIPD can offer such preferred pooled analysis. (v) Serving as the training set for other synthetic data methods: Our method can serve as the training set (or `pseudo IPD') for other generative models when the true IPD is unavailable, see Subsection \ref{app:pseudo IPD}.

The credibility of SynthIPD is demonstrated through two applications. Its applicability has in fact readily been verified by a surrogacy analysis between minimal residual disease and progression-free survival (PFS) in Multiple Myeloma (MM) \cite{ren2026leveraging}. Furthermore, the effectiveness of our method is demonstrated through two detailed case studies in this paper. The first case involves reconstructing IPD from the N9741 colorectal cancer trial \cite{goldberg2002n9741,sanoff2008five}, showing remarkable similarity with the actual trial data. In the second case study, synthetic IPD for six different MM trials are generated ($n=4061$) and a pooled analysis is conducted to validate the clinical benefit of Daratumumab in high-risk cytogenetic patients. Synthetic IPDs are then pooled together to establish benchmark estimates for this specific patient population.

Beyond case studies, simulation studies are conducted to further validate the accuracy and robustness of the method. Digitization accuracy is compared with the existing method IPDfromKM \cite{liu2021ipdfromkm}, showing obvious advantages of the proposed method. The accuracy of covariate generation is measured comparing synthetic versus truth using two distributional metrics. The distributions of SynthIPD closely align with the distributions of the true IPD, demonstrating the effectiveness of the proposed method. 

\subsection{Existing literature}\label{subsec:existing approach}
Several strategies have been developed with the aim of producing synthetic data sets. Two independent lines of work receive interest in the field. Our work falls into the second line but is closely connected to the first line as well, see Appendix \ref{app:pseudo IPD} for more discussions.

\textbf{Generative modelling.} First line of work generates synthetic data sets by training on real IPD, then synthetic data can be generated via `predicting' outcomes for a new test point. One of the most commonly seen approaches is to create data sets based on generative models \cite{azizi2021can,hahn2025generating,norcliffe2023survivalgan,brockschmidt2026survdiffdiffusionmodelgenerating,xu2019modeling}. Given the large number of similar methods, we refer readers to, e.g., \cite{PEZOULAS20242892} for a detailed reading therein.  The performance of trained generative models depends heavily on the quality of training data. Also, the utility of such approach is limited by access to real IPD, which is hard in clinical trial studies. Some other techniques attempt to impose distributional assumptions or leverage externally available population summaries \cite{pham2019population,gu2023synthetic}, but such imputations can introduce bias if the underlying assumptions are violated. 

\textbf{IPD reconstruction.} The other line of work focuses on reconstruction of IPD from published trials. The reconstruction can only recover what was reported and cannot generate more data like trained models. This can be done manually or with the assistance of large language models (LLMs) \cite{guyot2012enhanced,liu2021ipdfromkm,kmgpt}. Another IPD reconstruction method is KMsubtraction \cite{zhao2022kmsubtraction}. The key motivation of KMsubtraction is to recover subgroup IPD when only partial subgroup data are published. However, it requires access to both overall and subgroup KM plots; if these are unavailable, the method cannot be applied. Proposed methods in the literature have relatively lower reconstruction accuracy and cannot incorporate covariate structures. As real applications of IPD reconstruction methods, \cite{fell2021kmdata} constructs a curated data base purely based on IPD reconstruction for $304$ trials and \cite{syn2021association} publishes an article in \textit{The Lancet} utilizing mainly IPD reconstruction methods. 

\subsection{Our contributions}
Our contribution is threefold. First, we introduce an innovative methodological framework that reconstructs IPD, without requiring access to any true IPD. Second, unlike existing IPD reconstruction approaches \cite{liu2021ipdfromkm,guyot2012enhanced,kmgpt}, SynthIPD is the first to leverage SVG representations of KM plots, enabling high-precision recovery of event times and censoring patterns. Third, the core of our approach is an interpretable optimization procedure that enforces consistency with subgroup-level statistics, thereby facilitating downstream covariate-adjusted analyses. In summary, SynthIPD transforms published trial results into analyzable, high-fidelity synthetic datasets, creating a new foundation for methodological research that traditionally depends on, possibly inaccessible, IPD.

The paper is structured as follows: Section \ref{sec:conceptual framework} introduces the motivation and two meta-data sets used in our study. Section \ref{sec:proposed method} details statistical preliminaries and elaborates on the proposed method. Section \ref{sec:case studies} presents the results of the real-world applications. Section \ref{sec:sim} provides a very brief summary of the simulation studies conducted and the details are deferred to Appendix \ref{sec:simulation}. Section \ref{sec:conc} concludes future directions. Additionally, a comparison of our method with five generative models---survivalGAN \cite{norcliffe2023survivalgan}, CTGAN \cite{xu2019modeling}, TVAE \cite{xu2019modeling}, SurvDiff \cite{brockschmidt2026survdiffdiffusionmodelgenerating}, and conditional trees \cite{azizi2021can}---is given in Appendix \ref{app:pseudo IPD}.

\section{Data sets and scientific questions}\label{sec:conceptual framework}
Scientific progress often depends on analyses of high-quality data sets. In many areas, the fundamental problem is not only the quality of data, but also the accessibility of data. IPD from clinical trials, arguably among the most valuable data in medical research, often remain inaccessible due to privacy regulations, proprietary ownership, and logistical restrictions. 

In clinical trial practice, data for each randomized patient is collected following a prespecified protocol. The structure of IPD with survival endpoint typically consists of survival time, censoring status, treatment arms and covariates. However, when results are made public, only aggregate summaries, for example, the KM plot for all randomized patients and subgroup summary tables reporting event sizes, median survivals, and hazard ratios, are typically disclosed. Such conventional pipeline (See Figure \ref{fig:survival analysis} for a simple illustration)
$$\text{[I] IPD with covariates} \to \text{[II] KM curves, [III] Subgroup summaries}$$
is well established and can be implemented by standard survival-analysis tools, e.g., R package `\texttt{survival}'. The reverse process 
$$\text{[II] KM curves, [III] Subgroup summaries}\to \text{[I] IPD with covariates}$$
has long been regarded as infeasible. The goal of this work is to establish a framework for generating synthetic IPD, going from [II],[III] to [I].

Conventionally, a dataset consists of rows of observed responses and covariates, typically numerical, and data analysis involves fitting a model that predicts or explains the responses from the covariates. The structure of our problem is fundamentally different, we call our data sets \textbf{meta-data} sets. The `covariates' in a meta-data set are [II] and [III] in published trial reports, while [I] are the unobserved `responses'. The goal of SynthIPD is to predict [I] using [II], [III]. In this sense, the analyzed meta-data under our framework is \textbf{non-numerical} with structured graphical and tabular covariates and \textbf{unobservable responses}.

To demonstrate the motivating research problems, we consider two special datasets of high clinical importance.

\subsection{The first meta-data set (N9741)}

N9741 \cite{goldberg2002n9741,sanoff2008five} is a Phase III clinical trial comparing three chemotherapy regimens for previously untreated metastatic colorectal cancer. We are typically interested in two oxaliplatin-based arms out of three studied arms: FOLFOX (oxaliplatin+infusional 5-FU/leucovorin, $n=421$) and IROX (irinotecan+oxaliplatin, $n=383$). The primary endpoint was progression-free survival (PFS), with secondary endpoints including overall survival (OS).

\begin{figure}[!t]
  \centering
  \resizebox{\textwidth}{!}{%   scales the TikZ picture to \textwidth
  \begin{tikzpicture}[
    boxstyle/.style={
        draw=black!25, line width=0.6pt, rounded corners=2pt,
        fill=white, inner sep=6pt
    },
    circlabel/.style={
        circle, draw=#1, line width=1.1pt, fill=#1!18,
        minimum size=11mm, font=\bfseries\large, text=#1!55!black,
        drop shadow={shadow xshift=0.5pt, shadow yshift=-0.8pt,
                     opacity=0.25, fill=black}
    },
    thickarrow/.style={
        -{Stealth[length=5mm, width=6mm]},
        line width=6pt, draw=arrowNavy
    },
    feedbackarrow/.style={
        -{Stealth[length=4mm, width=5mm]},
        line width=2.5pt, draw=arrowNavy
    },
    midlabel/.style={
        font=\small\itshape\bfseries, text=arrowNavy
    }
]
 
    % ========= [I] IPD table (left) =========
    \node[boxstyle] (ipd) at (0,0) {\ipdtable};
 
    % ========= [II] KM plot (top right) =========
    \node[boxstyle, anchor=south west] (km)
        at ($(ipd.east)+(6.2cm, 0.4cm)$)
        {\includegraphics[width=12cm]{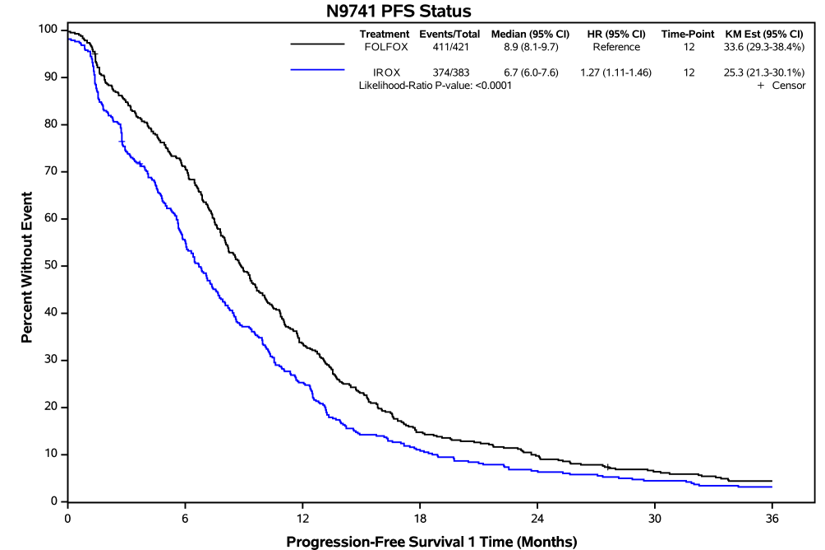}};
 
    % ========= [III] Subgroup table (bottom right) =========
    \node[boxstyle, anchor=north west] (sub)
        at ($(ipd.east)+(6.2cm,-0.4cm)$)
        {\subgrouptable};
 
    % ========= Two thick forward arrows from [I] to [II]/[III] =========
    \draw[thickarrow]
        ($(ipd.east)+(0.25, 0.9)$) -- ($(km.west)+(-0.25,-3.7)$);
    \draw[thickarrow]
        ($(ipd.east)+(0.25,-0.9)$) -- ($(sub.west)+(-0.25,2.72)$);
 
    % "Survival analysis" label between the arrows, with white halo
    \node[midlabel, fill=white, inner sep=3pt, rounded corners=1pt]
        at ($(ipd.east)+(3.2, 0)$) {\large Survival analysis};
 
    % ========= Curved feedback arrow [III] -> [I] with "?" on the curve =====
    \draw[feedbackarrow, dashed, dash pattern=on 5pt off 3pt]
        ($(sub.west)+(0,-2.15)$)
        .. controls +(0,-2.1) and +(0,-2.4) ..
        ($(ipd.south)+(1.2,-0.15)$)
        node[pos=0.5, font=\Huge\bfseries, text=arrowNavy,
             fill=white, inner sep=2pt, yshift=2pt] {?};
 
    % ========= Label circles at each panel's top-left =========
    \node[circlabel=labelTeal] at ([shift={(0.3, 0.35)}]ipd.north west) {I};
    \node[circlabel=labelNavy] at ([shift={(0.4, 0.40)}]km.north west)  {II};
    \node[circlabel=labelGold] at ([shift={(0.4, 0.40)}]sub.north west) {III};
 
\end{tikzpicture}
  }%
  \caption{An illustration by N9741 study. [I] Survival IPD collected per
    patient; [II] KM plot generated for all randomized patients; [III]
    subgroup summary statistics stratified by gender, including number of
    events, median progression-free survival, and hazard ratio. Both [II]
    and [III] are provided by Mayo Clinic. The dashed arc indicates the
    inverse problem addressed by SynthIPD: reconstructing [I] given [II]
    and [III] only.}
  \label{fig:survival analysis}
\end{figure}

The true IPD of N9741 is not available. However, the meta-data of N9741 study is available and provided. Specifically, the PFS KM plots for all randomized patients and the Gender subgroup analysis results for N9741 are presented in Figure \ref{fig:survival analysis} [II],[III]. These two pieces of information are crucial for the reconstruction of synthetic IPD.

% \begin{figure}[htbp]
%     \centering
%     \includegraphics[width=1\linewidth]{survival analysis.png}
%     \caption{An illustration by N9741 study: Survival IPD collected [I]; the KM plot generated for all randomized patients [II]; the subgroup summary statistics stratified by gender, including number of events, median progression-free survival, hazard ratio [III]. Both [II] and [III] are provided by Mayo Clinic.} 
%     \label{fig:survival analysis}
% \end{figure}

Aside from the clinical endpoints reported (median survival times, HRs) in N9741, many clinically important statistics routinely used in oncology, such as survival probabilities at 6 and 12 months, restricted mean survival time (RMST), etc., are not reported. Furthermore, although the biological heterogeneity of colorectal cancer has long been recognized, clinical trials, including N9741, are rarely performed to study the effect of a new intervention in patients with a specific biological subset. It is therefore of interest to discover unreported insights in N9741 and understand the performance of current treatments (FOLFOX and IROX) on a specific subgroup. 

\subsection{The second meta-data set (MM trials)}\label{subsec:questions}
\begin{figure}[htbp]
\centering
\begin{minipage}[b]{0.32\textwidth}
  \centering
  \includegraphics[width=\textwidth]{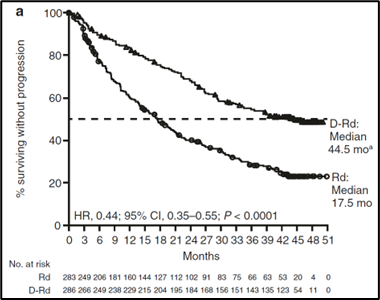}
\end{minipage}
\begin{minipage}[b]{0.32\textwidth}
  \centering
  \includegraphics[width=\textwidth]{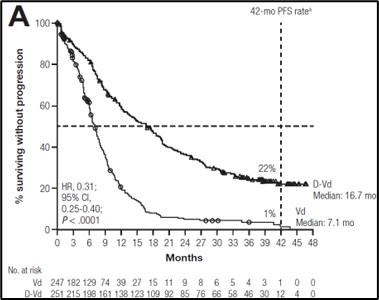}
\end{minipage}
\begin{minipage}[b]{0.32\textwidth}
  \centering
  \includegraphics[width=\textwidth]{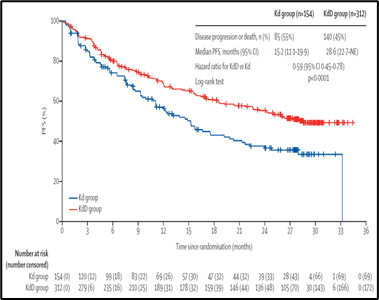}
\end{minipage}

\vspace{0.3cm}

\begin{minipage}[b]{0.32\textwidth}
  \centering
  \includegraphics[width=\textwidth]{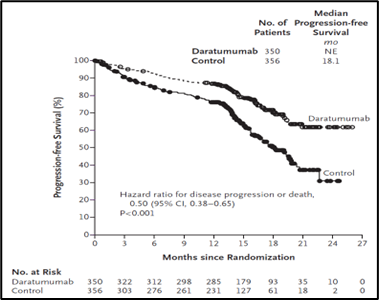}
\end{minipage}
\begin{minipage}[b]{0.32\textwidth}
  \centering
  \includegraphics[width=\textwidth]{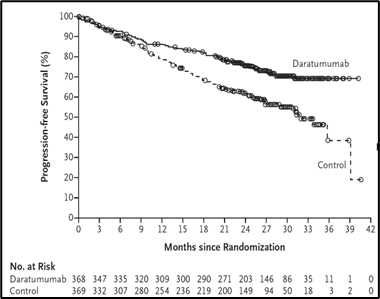}
\end{minipage}
\begin{minipage}[b]{0.32\textwidth}
  \centering
  \includegraphics[width=\textwidth]{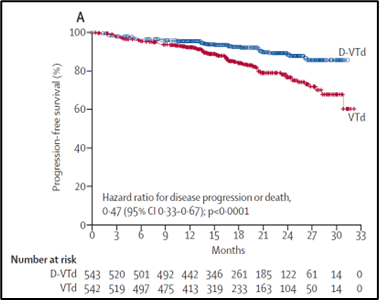}
\end{minipage}

\caption{The KM plots used to generate synthetic IPD. Top left: POLLUX ($n=286$ vs $283$), top middle: CASTOR ($n=251$ vs $247$), top right: CANDOR ($n=312$ vs $154$); bottom left: ALCYONE ($n=356$ vs $350$), bottom middle: MAIA ($n=369$ vs $368$), bottom right: CASSIOPEIA ($n=543$ vs $542$).}
\label{fig:6 subgroup MM}
\end{figure}

Six phase III trials that utilize Daratumumab-based regimens for MM are considered. Usually, Newly diagnosed MM patients (NDMM) may lead to better prognosis compared with relapsed/refractory MM (RRMM) patients. Thus, the $6$ screened trials are further separated into three focusing on NDMM: ALCYONE \cite{mateos2018daratumumab}, CASSIOPEIA \cite{moreau2019bortezomib}, MAIA \cite{facon2019daratumumab}; and the other three trials focusing on RRMM: CANDOR \cite{dimopoulos2020carfilzomib}, CASTOR \cite{palumbo2016daratumumab}, POLLUX \cite{dimopoulos2016daratumumab}. The primary endpoints for all $6$ studies are PFS. As usual, no IPD from any trial is given or analyzed in the paper. The meta-data of these trials are available: The PFS KM plots for all randomized patients and subgroup analysis results for cytogenetic risk profile are summarized and presented in Figure \ref{fig:6 subgroup MM} and Table \ref{tab:dara_implementation} respectively. Readers can also find these information in the original publications cited.

\begin{table}[ht!]
\caption{Reported subgroup analysis table for $6$ studies (by true IPD). HRCA: high-risk cytogenetic abnormalities. NE: not estimable because median PFS is not reached due to short follow-up time.}
\label{tab:dara_implementation}
\begin{center}
{%
    \fontsize{9pt}{11pt}\selectfont
\begin{tabular}{ccccccc}
&\multicolumn{2}{c}{Daratumumab} &&\multicolumn{2}{c}{Control}& HR ($95\%$ CI)\\
 \cline{2-3}\cline{5-6}
  Study names& Events/Total& mPFS ($95\%$ CI)&& Events/Total& mPFS ($95\%$ CI) & Dara vs Control\\\hline
  \textbf{ALCYONE} &&&&&&\\
  HRCAs &24/53&18.0&&19/45&18.1&0.78 (0.43--1.43)\\
  Standard &54/261&NE&&108/257&17.4&0.39 (0.28--0.55)\\
  \hline
  \textbf{MAIA} &&&&&&\\
   HRCAs&18/48&NE&&17/44&NE&0.85 (0.44--1.65)\\
  Standard &67/271&NE&&111/279&31.2&0.49 (0.36--0.67)\\
  \hline
  \textbf{CASSIOPEIA} &&&&&&\\
   HRCAs&15/82&-&&22/86&-&0.67 (0.35--1.30)\\
  Standard&30/460&-&&69/454&-&0.41 (0.26--0.62)\\
  \hline
  \textbf{POLLUX} &&&&&&\\
   HRCAs&23/35&26.8&&29/35&8.3&0.34 (0.16--0.72)\\
  Standard&87/193&NE&&124/176&18.6&0.43 (0.32--0.57)\\
  \hline
  \textbf{CASTOR} &&&&&&\\
   HRCAs&33/40&12.6&&31/35&6.2&0.41 (0.21--0.83)\\
  Standard&98/141&16.6&&117/140&6.6&0.26 (0.19--0.37)\\
  \hline
  \textbf{CANDOR} &&&&&&\\
   HRCAs&30/48&15.6 (7.6--26.2)&&18/26&5.6 (3.6--9.5)&0.49 (0.26--0.92)\\
  Standard&39/107&NE (25.9--NE)&&26/56&16.6 (11.1--NE)&0.54 (0.32--0.91)\\
  Unknown & 71/157 & 28.1 (18.6--NE) && 41/72 & 15.7 (11.1--28.0) & 0.64 (0.43--0.94)\\
  \hline
\end{tabular}
}
\end{center}
\end{table}

When planning a new trial, it is crucial to determine a specific patient population of interest and estimate the expected outcome (typically the median survival time) of the control arm in that population. Many studies, including the six trials considered here, investigate similar patient groups with comparable treatment arms. However, conventional aggregation techniques, such as meta-analysis, are not suited for median survival times because these statistics are neither asymptotically normal nor symmetric. In practice, median survival times are often inferred from clinical experience, real-world registries or meta-analytic estimates, all of which can introduce bias. Consequently, developing reliable methods to aggregate survival information, beginning with the six MM studies as a case study, is crucial for establishing a robust benchmark to inform future trial planning.

Summarizing all discussions above, an important and pressing research question would be: whether synthetic IPD can be generated with high fidelity when we have no information on the true IPD? If yes, then we are able to discover unreported insights, to understand the performance of current treatment on a specific subgroup and to establish benchmarks for different treatment options, etc. We address the above research question in Section \ref{sec:proposed method} by proposing a method to generate synthetic IPD. The analyses of these two meta-data sets are then discussed in Section \ref{sec:case studies}.

\section{Framework and methodology}\label{sec:proposed method}
To address the challenges posed by inaccessible IPD in the datasets above, we develop SynthIPD, an IPD reconstruction framework leveraging published survival summaries. The goal of SynthIPD is not the exact recovery of the original individual patient data. In general, multiple distinct IPDs may induce the same arm-level KM curves, at-risk tables and summary statistics. SynthIPD should be interpreted as one synthetic dataset from a class of IPDs that possess similar statistical properties with the true IPD and enable consistent downstream analyses.

\subsection{Preliminaries}\label{subsec:prelim}
Consider a clinical trial study of $n$ patients with survival outcome. For the $i$th patient, denote the observed survival time $U_i=\min(T_i, C_i)$, where $T_i$ is the time to disease progression or death, $C_i$ is the administrative censoring time and $\Delta_i=\mathbbm{1}\{T_i\leq C_i\}$ is the censoring status. Let $A_i$ be the treatment assignment, that is, $A_i=1$ if active treatment and $A_i=0$ if control. Set $X_i$ to be a univariate categorical covariate, taking values in $\{0,\ldots,K\}$. The multivariate covariates case is discussed in Subsection \ref{subsubsec:multivariate}. The (true) IPD from a trial with survival endpoint is defined as the collection $D_n:= \{U_i,\Delta_i,A_i,X_i\}_{i=1}^{n}$
with finite sample realization $d_n=\{u_i,\delta_i,a_i,x_i\}_{i=1}^{n}$. The goal is to generate a synthetic copy $$\tilde{d}_n :=\left\{\tilde{d}_n^{(-x)},\{\tilde{x}_i\}_{i=1}^{n}\right\} = \{\tilde{u}_i,\tilde{\delta}_i,\tilde{a}_i,\tilde{x}_i\}_{i=1}^{n}$$
with similar statistical properties as $d_n$. 

% \begin{remark}
%     The Assumption on univariate categorical covariate is a consequence of information available in practice. Most randomized controlled trials report routinely the summary statistics stratified by no more than two categorical factors together, see, e.g., EMA guidline on subgroup analyses \cite{ema2019subgroup}. Also, as reported in \cite[Table 2]{williamson2022subgroup}, $~94\%$ of the $178$ randomized controlled trials from $2016$ to $2021$ that utilized continuous information are dichotomized, and $66$ trials never used continuous information. 
% \end{remark}

\begin{remark}
    The assumption that $X_i$ is a univariate categorical covariate reflects the information typically available in published trial reports. Subgroup survival summaries are usually reported marginally, one factor at a time. For example, forest plots commonly present treatment effects within each level of each baseline covariate \cite{ruiwang2007reportingsubgroup}, and regulatory guidance also treats single-factor subgroup definitions as sufficient in many settings \cite{ema2019subgroup}. Hence, the summaries needed to reconstruct a joint covariate vector, namely summaries within every stratum and treatment arm, are not usually available. 
\end{remark}

For each treatment arm $a\in\{0,1\}$ and subgroup $x\in\{0,\ldots,K\}$, let $n_x$ (resp. $n_{x,a}$) be the number of patients in subgroup $x$ (resp. subgroup $x$, treatment $a$). Denote by
$$S_a(t):=P(T\geq t\mid A=a), \quad S_{x,a}(t) := P(T\geq t\mid A=a,X=x)$$
the survival functions stratified by treatment or treatment and covariate. The hazard function and conditional hazard function can be expressed as 
$$\lambda_a(t) = \lambda_0(t)\exp(\theta a),\quad \lambda_{x,a}(t) =\lambda_x(t)\exp(\theta_x a)$$
respectively, where $\lambda_0(t), \lambda_x(t)$ are the baseline hazard functions, $\theta,\theta_x$ are the treatment effects. Commonly used statistics describing the distribution of $T$ within subgroup $A=a,X=x$ includes the median survival time and conditional hazard ratio
$$m_{x,a} :=\inf\{t:S_{x,a}(t)\leq 0.5\},\quad \tHR_x:= \exp(\theta_x).$$
% Let $\ell_{m_{x,a}}, u_{m_{x,a}}$ (resp. $\ell_{\tHR_{x}},u_{\tHR_{x}}$) be the $95\%$ confidence intervals for $m_{x,a}$ (resp. $\tHR_x$). 
Given data $d_n$, finite-sample estimates for above statistics and their $95\%$ confidence intervals ($95\%$ CI) can be obtained by using standard survival analysis packages, e.g., R package `\texttt{survival}'. Let $g$ be a function and $\hat{g}$ its estimate based on observed data. The finite-sample estimates $\hat{m}_{x,a},\hat{\tHR}_x$ are essentially functions of the form
$$\hat{g}_x:d_{n,x}\mapsto \hat{g}_x(d_{n,x})\in\mathbb{R},\quad \hat{g}_{x,a}: d_{n,x,a}\mapsto \hat{g}_{x,a}(d_{n,x,a})\in\mathbb{R}$$
where $d_n$ is the IPD, $d_{n,x}$ (resp. $d_{n,x,a}$) is the IPD in subgroup $X=x$ (resp. in subgroup $X=x,A=a$). Further, for any type of estimate, we define the collection of summary statistics
$$
    \mathcal{J}_{\hat{g}}:=\left\{\text{$\hat{g}$, lower $95\%$ CI of $\hat{g}$, upper $95\%$ CI of $\hat{g}$}\right\},
$$
% For any statistic, say $g_x(F), F\in \mathcal{F}_x$, the finite sample estimator based on the observed sample $d_n$ will be denoted as $\hat{g}_x(d_n)$. The corresponding upper (resp. lower) confidence intervals constructed is denoted as $u_{\hat{g}_x}(d_n)$ (resp. $\ell_{\hat{g}_x}(d_{n,x})$). 
and adopt maximum Relative Absolute Error (mRAE) that penalize worst‐case discrepancies across all reported summary statistics:
\begin{equation}\label{eq:loss1}
    \mathcal{L}_m(\tilde{d}_n\mid d_n) := \max_{\hat{g}_{x,a}\in \mathcal{J}_{\hat{m}_{x,a}}}\left|\frac{\hat{g}_{x,a}(d_{n,x,a})-\hat{g}_{x,a}(\tilde{d}_{n,x,a})}{\hat{g}_{x,a}(d_{n,x,a})}\right|\in[0,1].
\end{equation}
\begin{equation}\label{eq:loss2}
    \mathcal{L}_{\tHR}(\tilde{d}_n\mid d_n) := \max_{\hat{g}_x\in \mathcal{J}_{\widehat{\tHR}_x}}\left|\frac{\hat{g}_x(d_{n,x})-\hat{g}_x(\tilde{d}_{n,x})}{\hat{g}_x(d_{n,x})}\right|\in[0,1].
\end{equation}
mRAEs \eqref{eq:loss1} and \eqref{eq:loss2} measure how much the summary statistics generated by synthetic data deviates away from the truth. For simplicity, we use $\mathcal{L}_{m},\mathcal{L}_{\tHR}$ if no confusion is introduced. In the best case, with $\tilde{d}_n=d_n$, the losses will be exactly $0$'s. These two losses will be a crucial quantity in our method and will be used as one of the similarity measures between two IPDs. 

\subsection{SynthIPD algorithm}\label{subsec:synthipd}

Suppose a clinical trial study with two treatment is of interest but its IPD is not disclosed. Instead, the following are published in, possibly, a list of peer-reviewed clinical trial articles: a KM survival curve for all randomized patients in treatment (resp. control) in Vector Graphics (VG) format along with a shared time grid $\mathcal{T}:=\{t_0,\ldots, t_{\ell}\}$ and an at risk number count $\mathcal{N}_1 :=\{N_{1,t_0},\ldots,N_{1,t_\ell}\}$ (resp. $\mathcal{N}_0 :=\{N_{0,t_0},\ldots,N_{0,t_\ell}\}$) and the summary statistics within subgroups $\hat{\mathcal{J}}_{\hat{m}_{x,a}},\hat{\mathcal{J}}_{\widehat{\tHR}_x}$ for $a=0,1$.  We assume that $\hat{\mathcal{J}}_{\hat{m}_{x,a}},\hat{\mathcal{J}}_{\widehat{\tHR}_x}$ are derived from the same population used to generate the KM plots, otherwise this piece of information cannot be used. In addition, we assume these graphical and tabular inputs should be coherent up to ordinary reporting imprecision such as rounding or plotting granularity. The risk table information $\{\mathcal{N}_a, \mathcal{T}:a\in\{0,1\}\}$ is crucial for the generation of survival IPD without covariate information while $\hat{\mathcal{J}}_{\hat{m}_{x,a}},\hat{\mathcal{J}}_{\widehat{\tHR}_x}$ are used to generate synthetic covariates.

\begin{remark}[Scope and required inputs] 
The KM plots and at risk numbers are usually reported together as Figures in a clinical trial paper (see, e.g., \cite[Figure 2]{dimopoulos2020carfilzomib}) while the subgroup information is usually reported in a separate forest plot (see, e.g., \cite[Figure 3]{dimopoulos2020carfilzomib}). These are practical assumptions and are routinely reported in current trials using survival as endpoints \cite[Table 1]{ou2020guidelines}, \cite[Section B]{fda2025survival}. For a more detailed illustration, see Table \ref{tab:data-requirements} in Appendix \ref{sec:practical requirements}. 
\end{remark}

\begin{figure}[t!]
    \centering
    \includegraphics[width=\linewidth,scale=1]{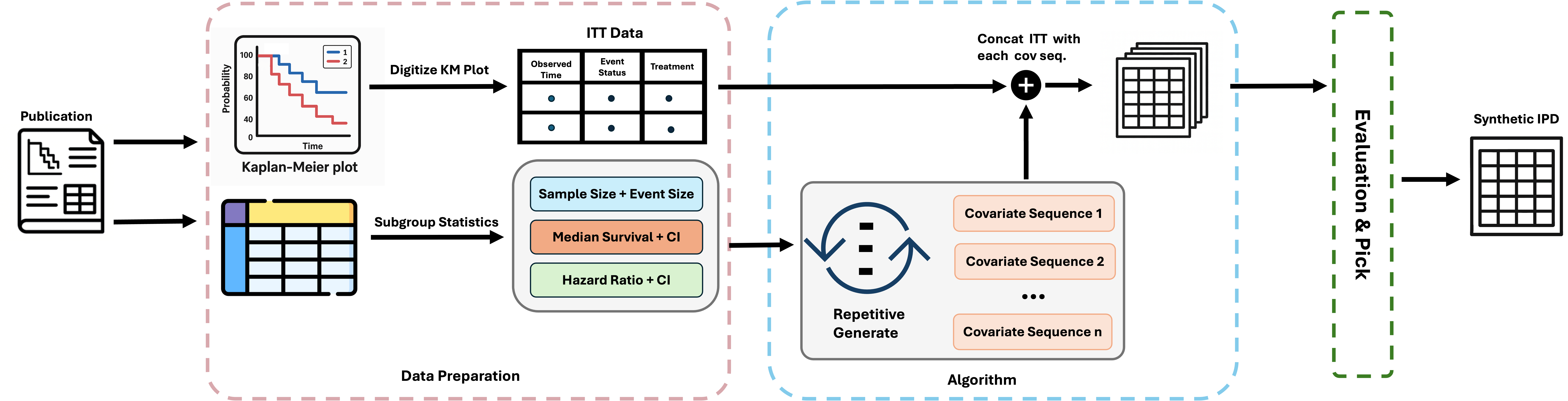}
    \caption{The workflow of SynthIPD.}
    \label{fig:flowchart}
\end{figure}

The SynthIPD method can be formalized into a three-step procedure: 

\noindent Step 1. a digitization step (DIGITIZE($\mathcal{N}_a,\mathcal{T}:a\in\{0,1\}$)) producing $\tilde{d}_n^{(-x)}$. See Figure \ref{fig:flowchart} Data Preparation block.

\noindent Step 2. a covariate-generation step (Cov-Generation) producing $\{\tilde{x}_i\}_{i=1}^{n}$, where $R,H,\alpha,\beta,f,S_m,S_{\tHR}$ are hyper-parameters. See Figure \ref{fig:flowchart} Algorithm block.

\noindent Step 3. a combination step which combines the data in Step 1 and 2 into $\tilde{d}_n$. The third step is straight forward and requires no further explanation. See Figure \ref{fig:flowchart} Evaluation\&pick block.

\begin{table}[t!]
    \centering
    \begin{tabular}{llll}
    Step & Input & Hyper-parameters & Output \\\hline
    DIGITIZE & $\mathcal{N}_1,\mathcal{N}_0,\mathcal{T}$ & None &$\tilde{d}_{n}^{(-x)}$ \\
    Cov-Generation & $\tilde{d}_n^{(-x)}, \hat{\mathcal{J}}_{\hat{m}_{x,a}},\hat{\mathcal{J}}_{\widehat{\tHR}_x}$ & $R,H,\alpha,\beta,f,S_m,S_{\tHR}$ &$\{\tilde{x}_i\}_{i=1}^n$ \\
    Combine & $\tilde{d}_n^{(-x)},\{\tilde{x}_i\}_{i=1}^n$ & None &$\tilde{d}_n$ 
    \\\hline
    \end{tabular}
    \caption{Overview of inputs and outputs for each step.}
    \label{tab:Overview of outputs}
\end{table}

To clarify, Table \ref{tab:Overview of outputs} is presented with inputs and outputs for each step. The implementation of the proposed algorithm requires several practical requirements, which is not stringent and can be satisfied by the publications in most high quality clinical journal publications. An illustrative example of our workflow can be found in Example \ref{example:Candor}. 

\subsubsection{DIGITIZE}

Each drop (resp. tick mark) in the KM curve corresponds to a distinct event (resp. censoring). The VG figures are first transformed into SVG using tools like Inkscape, then exact coordinates of drops and tick marks can be retrieved using software tools like Figma, giving $\{\tilde{u}_i,\hat{S}(\tilde{u_i})\}_{i=1}^{m},m\leq n$. Additionally, information on $\tilde{a}_i,\tilde{\delta}_i,i\in[m]$ are available from the graph directly. If $m=n$, then all extracted coordinates are unique and the digitization is readily done. In practice, usually $m\leq n$ and one drop (resp. censoring tick mark) may represent multiple events (resp. censorings). In this case, it is necessary to estimate how many patients experienced event (resp. are censored) when they share the same visual drop (resp. censoring tick mark). The procedure is detailed as follows.

For one treatment arm and one time grid $[t_\iota,t_{\iota+1})$, let
$s_{\iota1}<\ldots<s_{\iota J_\iota}, q_{\iota1}<\ldots<q_{\iota K_\iota}$ be the distinct extracted event-drop times and censoring times, respectively. Let 
$\rho_{\iota j}:=\hat{S}(s_{\iota j})/\hat{S}(s_{\iota j}^-)$
denote the KM step ratio extracted from the plot at event time $s_{\iota j}$, with $\hat{S}(t^-)$ being the left hand limit of $\hat{S}(t)$. Define $d_{\iota j}$ (resp. $c_{\iota k}$) to be the numbers of failures and censorings occurring at $s_{\iota j}$ (resp. $q_{\iota k}$). The number at risk before the first event in any time window can be recursively derived as 
$$Y_{\iota 1} = N_{t_\iota} - \sum_{q_{\iota k}<s_{\iota 1}}c_{\iota k}, \quad Y_{\iota(j+1)} = Y_{\iota j}-d_{\iota j}-\sum_{s_{\iota j}<q_{\iota k}<s_{\iota (j+1)}}c_{\iota k}, \quad j=1,\ldots,J_\iota-1.$$
In the ideal case, if the coordinates, $s_{\iota j}, q_{\iota k}$ extracted, are internally consistent (does not suffer from any internal bias like rounding, figure quality, etc.), the system admits an integer solution satisfying the following system of equations
$$\begin{cases}
    \sum_{j=1}^{J_{\iota}}d_{\iota j}+\sum_{k=1}^{K_\iota}c_{\iota k} = N_{t_\iota}-N_{t_{\iota+1}},\\
    1-\frac{d_{\iota j}}{Y_{\iota j}} = \rho_{\iota j}, & j=1,\ldots, J_{\iota}.
\end{cases}$$
The above holds by the calculation of KM estimate which motivates the following Algorithm \ref{alg:exact KM}. 

\begin{algorithm}[htbp]
\caption{DIGITIZE algorithm via optimization}\label{alg:exact KM}
\begin{algorithmic}
    \Require For fixed time grid $[t_{\iota},t_{\iota+1})$, observe unique drop coordinates $(s_{\iota 1},\ldots, s_{\iota J_\iota}),$ unique censor coordinates $(q_{\iota 1},\ldots, q_{\iota K_\iota})$, KM ratios $(\rho_{\iota 1},\ldots, \rho_{\iota J_\iota})$. Observe the total number of events $D_\iota$ and the total number of censoring $C_\iota$ in this grid.

    \For {$\iota = 1$ to $\ell$}
    \State Estimate $\{(d_{\iota j},c_{\iota k}):j=1,\ldots,J_\iota, k=1,\ldots, K_\iota\}$ by solving the following constraint optimization problem
    \begin{align}
        (d_{\iota},c_{\iota}) &= \argmin_{(d_{\iota },c_{\iota})\in \mathbb{Z}_+^{J_\iota}\times \mathbb{Z}_+^{K_\iota}}\sum_{j=1}^{J_{\iota}}\left(\rho_{\iota j}-\left(1-\frac{d_{\iota j}}{Y_{\iota j}}\right)\right)^2+\lambda\sum_{k=1}^{K_{\iota}}\left(c_{\iota k} - \frac{C_\iota}{K_\iota}\right)^2\nonumber\\
        &\text{subject to }\sum_{j=1}^{J_\iota} d_{\iota j} = D_\iota, \sum_{k=1}^{K_\iota}c_{\iota k} = C_\iota\label{eq:exact KM}
    \end{align}
    \EndFor
\end{algorithmic}
\end{algorithm}

\begin{remark}
    The optimization problem \eqref{eq:exact KM} minimize the cumulative risk of KM ratio conditional on the inputs. Algorithm \ref{alg:exact KM} consists of independent finite-dimensional integer optimization problems for each time grid, admits at least one solution which can be solved quickly but need not be unique. If the information on censorings $C_\iota$ is not reported in the risk table, then the algorithm reduces to unconstrained optimization, cannot guarantee the number of censorings/failures to match and will be less stable. See, e.g., Table \ref{tab:data-requirements}.

    The choice of target average $C_\iota/K_\iota$ for censoring shrinks the solution toward roughly uniform censoring over the $K_{\iota}$ visible censoring times, which mainly affects within-window placement of censorings and usually has little impact on the reconstructed KM curve once the event counts are already matched.
\end{remark}

The optimization \eqref{eq:exact KM} is similar in spirit to a quadratic programming approach proposed \cite[Section 2]{titman2026qpprogramming}. We also provide uncertainty quantification for internal inconsistencies. Compared with the IPD reconstruction method proposed in, e.g., \cite{guyot2012enhanced,liu2021ipdfromkm}, we believe the revised exact-KM method is more suitable for at least three reasons. First, it uses the exact product-limit structure of the Kaplan–Meier estimator. Second, failures and censorings are estimated jointly, not by post-hoc adjustment like \cite{liu2021ipdfromkm}. Third, the proposed method yields clean uncertainty quantification (See Theorem \ref{theorem:uncertainty for DIGITIZE} in Appendix \ref{app:uncertainty quantification DIGITZE}) that explains why such method is more accurate than conventional IPD reconstruction methods.

\begin{remark}
    SVG is utilized but not required in the first step, instead, we only require Vector Graphics (VG) figures. VG figures can be transformed into SVG using computer tools like Inkscape. The requirement of VG is less stringent than SVG. An investigation is made on the author's guideline for some high-impact medical journals including but not limited to New England Journal of Medicine, JAMA, The Lancet and Journal of clinical oncology. All journals explicitly state that they ``create vector drawings'' in published articles and request ``vector files/graphics'' file for artwork, which implies VG figures are available most of the time. Even when the VG requirement is not satisfied, the digitization step can be done by web plot digitizer (e.g., \url{https://automeris.io}) with IPDfromKM package \cite{liu2021ipdfromkm} or KM-GPT \cite{kmgpt}, sacrificing digitization precision, see discussions in Subsection \ref{subsec:digitization numerical} and Appendix \ref{app:uncertainty quantification DIGITZE} . 
\end{remark}

\subsubsection{Cov-Generation}\label{subsubsec:cov-gen}
After finishing the digitization step, we obtain $\tilde{d}_n^{(-x)}$. The generation step aims to create a vector $\{\tilde{x}_i\}_{i=1}^{n}$ such that the statistical properties of the combined synthetic sample $\tilde{d}_n$ will be similar to that of $d_n$. 

One way is to generate the covariate vector repeatedly from \resizebox{0.27\textwidth}{!}{$\text{Multinomial}(n,(n_0,\ldots, n_K)/n)$} distribution and select the best covariate vector possible according to some criteria, e.g., the vector with minimal $\mathcal{L}_m,\mathcal{L}_\tHR$. This approach is natural because $n_x/n$ is guaranteed to approximate the distribution of $X$ well for large sample. However, it has two major drawbacks: First, it is computationally intensive to traverse all possible vectors $\{0,\ldots,K\}^n$ as $n,K$ increases. Secondly, it does not make use of the information $n_{x,a}$ provided in the subgroup analysis table. To improve upon these two drawbacks, we propose the Cov-Generation($\tilde{d}_n^{(-x)},\hat{\mathcal{J}}_{\hat{m}_{x,a}},\hat{\mathcal{J}}_{\widehat{\tHR}_x};H,R,\alpha,\beta,f,S_m,S_\tHR$) Algorithm \ref{alg:algorithm}. 

\begin{algorithm}[t!]
\caption{Cov-Generation($\tilde{d}_n^{(-x)},\hat{\mathcal{J}}_{\hat{m}_{x,a}},\hat{\mathcal{J}}_{\widehat{\tHR}_x};H,R,\alpha,\beta,f,S_m,S_{\tHR}$)}\label{alg:algorithm}
\begin{algorithmic}
\Require IPD without covariates $\tilde{d}_n^{(-x)}$, reported summary statistics $\hat{\mathcal{J}}$, 
max number of loss updates $H$, max number of random switches $R$, perturbation rate $\mathcal{C}\sim\mathrm{Unif}(\alpha,\beta)$,
acceptance probability function $f$, termination thresholds $S_m,S_{\tHR}$.
\State \textbf{Initialize:} $\tilde{x} \in \{0,\ldots,K\}^n$ subject to $\tilde n_{x,a} = n_{x,a}, \forall x, \forall a$
\State Form $\tilde{d}_n \gets \{\tilde{u}_i,\tilde{\delta}_i,\tilde{a}_i,\tilde{x}_i\}_{i=1}^n$
\State $(\mathcal{L}_m, \mathcal{L}_{\tHR}) \gets (+\infty, +\infty)$
\For{$h = 1$ to $H$}\Comment{Outer iteration}
    \For{$r = 1$ to $R$}\Comment{Inner iteration}
        \State Draw $\mathcal{C} \sim \mathrm{Unif}(\alpha,\beta)$ and propose $\tilde{x}'$ by switching $\mathcal{C}\%$ entries of $\tilde{x}$
        \State $\tilde{d}_n' \gets \{\tilde{u}_i,\tilde{\delta}_i,\tilde{a}_i,\tilde{x}'_i\}_{i=1}^n$
        \State $\mathcal{L}_m', \mathcal{L}_{\tHR}' \gets \mathcal{L}_m(\tilde{d}_n' \mid d_n),\; \mathcal{L}_{\tHR}(\tilde{d}_n' \mid d_n)$
        \State $\Delta \gets (\mathcal{L}_m' - \mathcal{L}_m)/\mathcal{L}_m$ \Comment{relative loss change}
        \If{$\mathcal{L}_m' < \mathcal{L}_m \ \mathbf{and}\  \mathcal{L}_{\tHR}' < \mathcal{L}_{\tHR}$} \Comment{strict improvement}
            \State $(\tilde{d}_n,\tilde{x},\mathcal{L}_m,\mathcal{L}_{\tHR}) \gets (\tilde{d}_n',\tilde{x}',\mathcal{L}_m',\mathcal{L}_{\tHR}')$; \textbf{break}
        \ElsIf{$\mathcal{L}_{\tHR}' < \mathcal{L}_{\tHR}$ \textbf{and} \text{Bernoulli}$\big(f(\Delta;r)\big)=1$} \Comment{Exploration}
            \State $(\tilde{d}_n,\tilde{x},\mathcal{L}_m,\mathcal{L}_{\tHR}) \gets (\tilde{d}_n',\tilde{x}',\mathcal{L}_m',\mathcal{L}_{\tHR}')$; \textbf{break}
        \EndIf
    \EndFor
    \If{$\mathcal{L}_m < S_m$ \textbf{and} $\mathcal{L}_{\tHR} < S_{\tHR}$} \textbf{break} \Comment{targets achieved}\EndIf
\EndFor
\State \Return $\tilde{d}_n$
\end{algorithmic}
\end{algorithm}

Algorithm \ref{alg:algorithm} takes the digitized data and iteratively updates the subgroup vector to minimize the median- and HR-based RAE losses \eqref{eq:loss1}, \eqref{eq:loss2}. The outer loop (indexed by $h$, up to $H$) counts accepted loss updates and the inner loop (indexed by $r$, up to $R$) counts number of proposals within each update. $\alpha,\beta$ represent the proportion of entries perturbed per proposal, and the acceptance function $f:(0,\infty)\times\mathbb{N}\to (0,1)$ satisfies $\lim_{\Delta\to\infty}f(\Delta,\cdot)= 0$, so that exploration is allowed while moves that substantially worsen the RAE loss are rejected with probability approaching one (see Subsection \ref{subsec:opt} for the construction of $f$). Recommended defaults are $H\approx 50\text{--}100$, $R\approx 5\times10^5\text{--}10^6$, $\alpha\geq 1$, $\beta\leq 20$, $S_m\approx 2\%$, $S_{\tHR}\approx 1\%$. The quality depends on the fidelity of the digitized input, and $\mathcal{L}_{\tHR}$ is generally easier to control than $\mathcal{L}_m$. A larger $\beta-\alpha$ perturbs more entries per step and stabilizes more slowly; adaptively shrinking $\alpha,\beta$ as the loss decreases is left to future work. A sensitivity analysis for different hyper-parameter combinations is given in Appendix \ref{app:sensitivity}.

As a running example, an example SynthIPD workflow implemented on one trial is provided in Example \ref{example:Candor}.

\begin{example}[Candor study \cite{usmani2022carfilzomib}]\label{example:Candor}
   The CANDOR study compares the use of carfilzomib, daratumumab, and dexamethasone (KdD) versus carfilzomib and dexamethasone (Kd) in relapsed or refractory MM patients. The primary endpoint is PFS. The KM plot and the subgroup summary statistics are provided in Figure \ref{fig:candor_summary} as inputs. The output of each step of our workflow is presented on the right. 

    \begin{figure}[htpb!]
        \centering
        \includegraphics[width=0.85\linewidth]{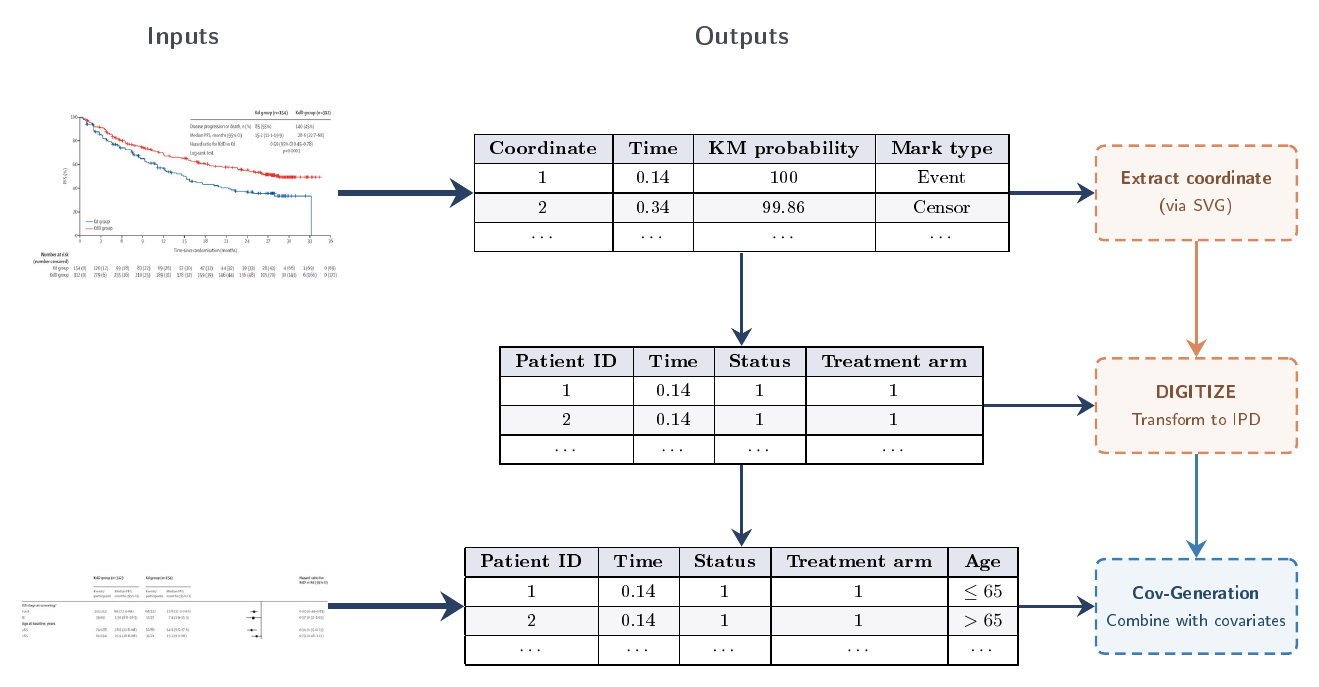}
        \caption{A running example of the method on CANDOR study. The inputs are: Reported KM curves taken from \cite[Figure 2]{usmani2022carfilzomib} and reported summary statistics taken from \cite[Figure 3]{usmani2022carfilzomib}. The outputs are: extracted coordinates by SVG; transformed IPD by DIGITIZE; IPD with covariate by Cov-Generation.}
        \label{fig:candor_summary}
    \end{figure}
\end{example}

\subsubsection{Extension to multivariate case}\label{subsubsec:multivariate}

Suppose multiple categorical covariates are available and $\vX_i = (X_{i1},\ldots, X_{ip})\in \{0,\ldots,K_1\}\times\ldots\times\{0,\ldots,K_p\}$. The goal is to reconstruct synthetic data $\tilde{d}_n=\{\tilde{d}_n^{(-x)},\tilde{\bm{x}}_i\}$. Denote by $S_i = S(\vX_i)\in [\prod_{j=1}^{p}(K_j+1)]$ a one-to-one stratum mapping. The covariate $S_i$ is then univariate with $\prod_{j=1}^{p}(K_j+1)$ levels. If $\hat{\mathcal{J}}_{\hat{m}_{s,a}}$ (resp. $\hat{\mathcal{J}}_{\widehat{\tHR}_{s}}$) the median survival (resp. HR) within each stratum $s$ and each treatment arm $a$ are reported, then the same algorithm in Subsection \ref{subsubsec:cov-gen} can be extended to the multivariate case. Note that for multivariate cases, the summary statistics within each stratum $s$ are needed. The marginal summaries are not enough to recover a full set of synthetic covariates. While the extension to multivariate covariates is useful and plausible, the joint reconstruction may be limited by data availability since not many clinical trials report results within strata of more than $2$ categorical covariates. Without strata information, modeling the covariates' covariance structure may improve the reconstruction. This is left for future research.

\subsection{Optimization, simulated annealing and interpretations}\label{subsec:opt}
We now state the problem as a complex constrained optimization problem and relate Algorithm \ref{alg:algorithm} to the SA algorithm. Given $\tilde{d}_n^{(-x)},d_n$, The Cov-Generation method essentially solves the following constrained optimization problem:
\begin{align}
    \tilde{\bm{x}} 
    &= \text{argmin}_{\bm{x}\in \{0,\ldots,K\}^{n}} 
    \Big(
    \mathcal{L}_m\!\left\{(\tilde{d}_n^{(-x)},\bm{x}) \mid d_n\right\},
    \mathcal{L}_{\tHR}\!\left\{(\tilde{d}_n^{(-x)},\bm{x}) \mid d_n\right\}
    \Big), \nonumber\\
    &\quad \text{subject to } 
    \sum_{i=1}^{n}\mathbbm{1}\{x_i=x,\,a_i=a\} = n_{x,a}, 
    \quad \forall\, x,a. 
    \label{eq:constrained_optimization}
\end{align}
The optimization problem in Equation \eqref{eq:constrained_optimization} is a discrete, non-convex, multi-objective optimization program that does not admit a closed-form solution. 
The global optimum under the multi-objective framework, denoted by \(\bm{x}^\star\), is defined as the covariate vector that simultaneously satisfies
\[
\mathcal{L}_m\!\left\{(\tilde{d}_n^{(-x)},\bm{x}^\star) \mid d_n\right\} 
\leq 
\mathcal{L}_m\!\left\{(\tilde{d}_n^{(-x)},\bm{x}) \mid d_n\right\},
\quad 
\forall\, \bm{x},
\]
and
\[
\mathcal{L}_{\tHR}\!\left\{(\tilde{d}_n^{(-x)},\bm{x}^\star) \mid d_n\right\} 
\leq 
\mathcal{L}_{\tHR}\!\left\{(\tilde{d}_n^{(-x)},\bm{x}) \mid d_n\right\},
\quad 
\forall\, \bm{x}.
\]

Simulated annealing (SA) is a probabilistic technique for approximating the global optimum of a given function, typically used for global optimization problems whose objective function is intricate and can only be evaluated via costly function evaluations~\cite{simulatedannealing, bertsimas1993simulated}.

We first describe the SA formulation before relating it to our method. Let $r$ index the current proposal attempt, $i^{(r)}$ the current state of the solution, and $S_{i^{(r)}}$ the neighborhood of $i^{(r)}$. Let $t^{(r)}$ be the value of a temperature parameter, so named for the algorithm's close relation to the physical annealing of materials~\cite[Section~2.3]{delahaye2018simulated}. The SA algorithm aims to minimize a pre-specified loss function $h(\cdot)$, starting from an initial state $i^{(0)}$. At the $r$th proposal attempt ($r \geq 1$), it generates a candidate $j^{(r)} \in S_{i^{(r)}}$, compares $h(j^{(r)})$ with $h(i^{(r)})$, and either moves to $j^{(r)}$ (accepting the proposal) or stays at $i^{(r)}$ (rejecting the proposal), with acceptance probability
\begin{equation}\label{eq:sa prob}
    P(\text{accept }j^{(r)}) = \begin{cases}
    1, & \text{if } h(j^{(r)}) < h(i^{(r)}), \\
    \exp\!\left(\dfrac{h(i^{(r)}) - h(j^{(r)})}{t^{(r)}}\right), & \text{otherwise,}
\end{cases}
\end{equation}
where $\lim_{r \to \infty} t^{(r)} = 0$. The decreasing sequence $\{t^{(r)}\}$ is called `temperature' and the rate at which it decreases is called `cooling schedule'. For sufficiently small $t^{(r)}$, the system increasingly favors moves that reduce the loss and avoids states that increase it. With $t^{(r)} \equiv 0$, the procedure reduces to a greedy algorithm that makes only loss-decreasing transitions. Rather than restricting the acceptance rule to the exponential form in Equation~\eqref{eq:sa prob}, we allow a more general function $f$ in our method; we refer to~\cite{delahaye2018simulated, zheng2024use} for a more detailed treatment of SA.

If \(f(\cdot, \cdot) \equiv 0\) in Algorithm~\ref{alg:algorithm}, the Cov-Generation procedure reduces to a randomized local search~\cite[Algorithm~1]{zheng2024use}. Another possible choice of $f$ here is
\begin{equation}\label{eq:f-realization}
f(\Delta; r) = \exp\!\left(-\frac{\Delta}{T(r)}\right), \qquad T(r) = \frac{0.02}{1 + \exp\!\big(-(r - 3000)/2000\big)},
\end{equation}
where $\Delta$ is defined in Algorithm~\ref{alg:algorithm}. Intuitively, $f$ is used to ensure the algorithm will not be trapped at local minimum and exploits when $r$ is large. A detailed discussion on the calibration of $T(r)$ is provided in Appendix \ref{app:calibration SA}.

Similar to SA, we adopt the principle of exploring the solution space by traversing neighborhoods and occasionally accepting worsening solutions with a non-zero probability. The substantive distinction is that the relative loss $\Delta$ partially defines the temperature schedule. To see this, observe that
$$
    -\frac{\Delta}{T(r)} \;=\; -\left(\mathcal{L}_m' - \mathcal{L}_m\right)\cdot \frac{1}{\mathcal{L}_mT(r)}
$$
so the factor $\mathcal{L}_mT(r)$ in the denominator acts as the temperature. Since each outer iteration accepts at least one update within the inner iteration, $\mathcal{L}_m$ decreases across outer iterations $h$, so this factor plays the role of the classical cooling schedule. The factor $T(r)$ allows exploration that is reset at each outer iteration: it remains small when strict improvement is found quickly and rises only when repeated rejections signal a locally-difficult configuration. Consequently, Cov-Generation can be viewed as an SA algorithm whose temperature schedule is factored into an outer-iteration component ($\mathcal{L}_m$) and an inner-iteration component ($T(r)$), with relaxed definitions of the neighborhood, temperature, and transition probability functions.

To implement the SA algorithm on \(\mathcal{L}_m\) under our setup, we define the neighborhood of \(\bm{x}\) as \(\bm{x}' \in \{0, \ldots, K\}^n\) such that at least \(\alpha\%\) and at most \(\beta\%\) of the coordinates within the same subgroup class are swapped. The coordinates are swapped with probability 1 if both the median and HR losses decrease. Note that \(\mathcal{L}_{\tHR}\) can be controlled more easily than \(\mathcal{L}_m\). Therefore, we require a reduction in \(\mathcal{L}_{\tHR}\) at each update and apply an SA procedure on \(\mathcal{L}_m\). Specifically, given a reduction in \(\mathcal{L}_{\tHR}\), we may accept solutions that yield a slightly higher \(\mathcal{L}_m\) value with low probability.

{The optimization problem in Equation \eqref{eq:constrained_optimization} minimizes $\mathcal{L}_m$ and $\mathcal{L}_{\tHR}$ and subject to a constraint over the number of subjects. In survival analysis, median survival typically controls the event/censoring behavior. The HR describes the difference between two survival curves (treatment vs control). Our motivation originates from the simultaneous control over the loss of median survival, HR, together with CI's. Such control guarantees the behavior of the two KM curves and the synthetic data to be similar to the original data. }

\section{Case studies}\label{sec:case studies}
% In this section, we formally provide answers and results to the scientific questions proposed in Subsection \ref{subsec:questions}.

\subsection{SynthIPD for N9741 trial}\label{subsec:n9741}
To discover unreported insights, synthetic IPD based on Figure \ref{fig:survival analysis} [II] and [III] is generated. Then, a secondary analysis is carried out on the gender subgroup using the synthetic data. 

\begin{figure}[htbp]
  \centering
\includegraphics[width=\linewidth,height=0.5\textheight]{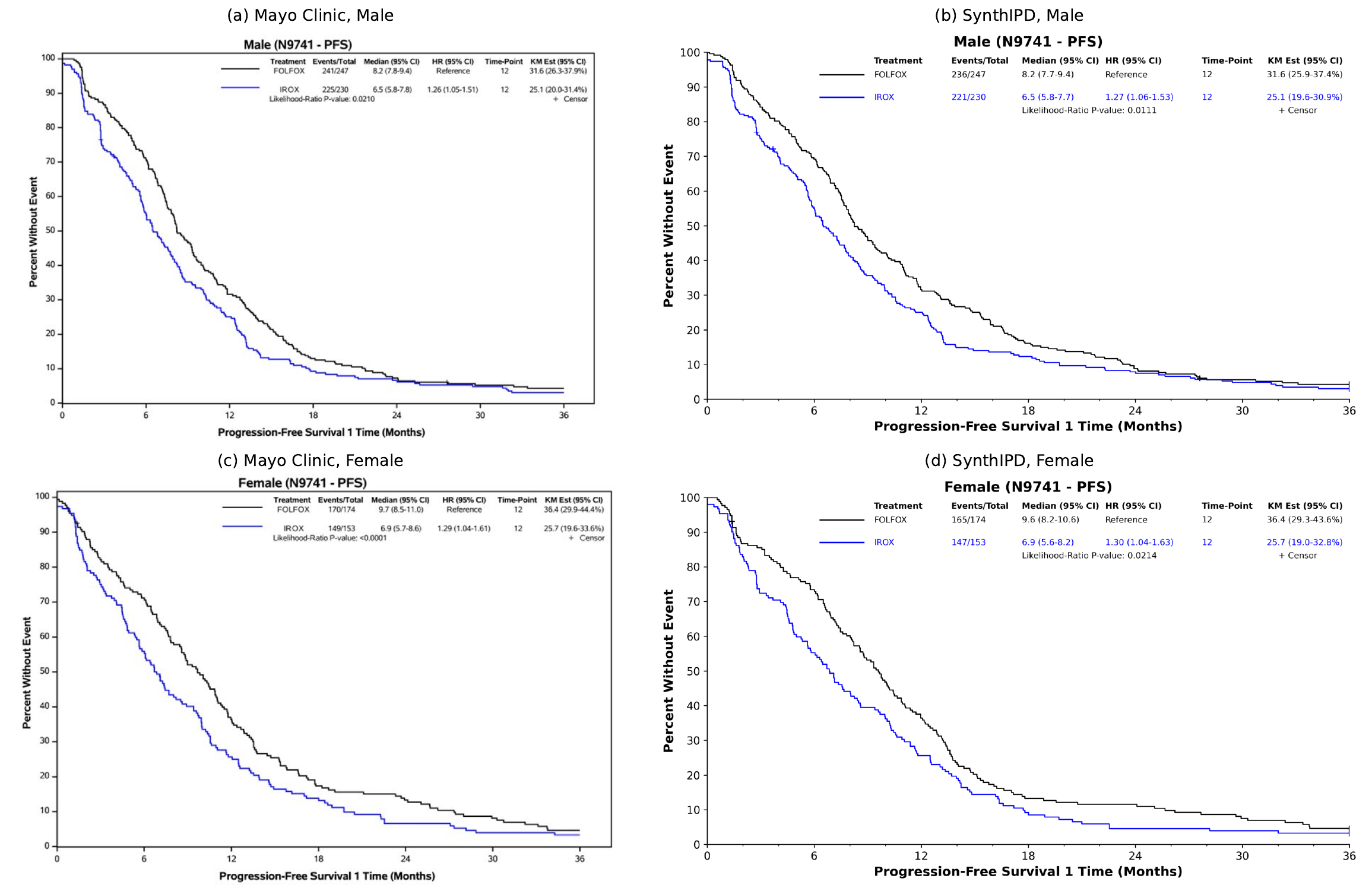}
  % % Row 1
  % \begin{minipage}[t]{0.48\textwidth}
  %   \centering
  %   \includegraphics[width=\linewidth,height=0.18\textheight]{n9741_male_mayo.png}
  %       \textbf{(a)}
  % \end{minipage}\hfill
  % \begin{minipage}[t]{0.48\textwidth}
  %   \centering
  %   \includegraphics[width=\linewidth,height=0.18\textheight]{n9741_male_synthetic.png}
  %   \textbf{(b)}
  % \end{minipage}

  % \vspace{0.3em}

  % % Row 2
  % \begin{minipage}[t]{0.48\textwidth}
  %   \centering
  %   \includegraphics[width=\linewidth,height=0.18\textheight]{n9741_female_mayo.png}
  %   \textbf{(c)}
  % \end{minipage}\hfill
  % \begin{minipage}[t]{0.48\textwidth}
  %   \centering
  %   \includegraphics[width=\linewidth,height=0.18\textheight]{n9741_female_synthetic.png}
  %   \textbf{(d)}
  % \end{minipage}

  % \vspace{0.3em}
  \caption{The subgroup KM plots (a)-(d) for N9741 study. The left $2$ plots are provided by Mayo Clinic for comparison with SynthIPD results only, the right $2$ plots are generated by SynthIPD. Two plots in the same row correspond to the same subgroup. }
  \label{fig:n9741_subgroup}
\end{figure}

The KM plot for all randomized patients (Figure \ref{fig:survival analysis} [II]) is digitized using DIGITIZE algorithm, then synthetic gender covariates are generated by Cov-Generation with parameters $H=100,R=10^6,\alpha=1,\beta=20,f\equiv 0,S_m=S_\tHR = 0$. The subgroup analysis results based on the true IPD and SynthIPD are shown in Table \ref{tab:n9741 synthIPD}. The results are identical. Figure \ref{fig:n9741_subgroup} further reports the KM curves within the gender subgroup. The KM curves of the synthetic IPD alignes closely with the KM curves provided by Mayo Clinic directly (for comparison only). 

The survival rate at a fixed time point represents the probability of surviving beyond that time. The RMST, defined as the area under the survival curve up to a prespecified truncation time, summarizes the average survival duration within a clinically meaningful window and remains valid even when the proportional hazards assumption does not hold. Both metrics capture treatment efficacy, complementing the relative comparison provided by the median survivals or HRs. However, these two clinically important metrics are not reported in the original publication. With the help of synthetic IPD, we are able to report these two metrics at months 12, 24, 36 in Table \ref{tab:new insights n9741}. In the N9741 trial, FOLFOX consistently demonstrated higher survival probabilities and longer RMST values than IROX across all evaluated time points for both female and male patients. The similarity of treatment contrasts across sexes suggests that the survival benefit of FOLFOX is not materially modified by gender. Results for another subgroup, age, are reported in Appendix \ref{app:additional info}. Potentially, synthetic data for all subgroups can be generated similarly, as long as the summary statistics are reported. We pick gender and age only as representative examples.

\begin{table}[]
    \centering
    {%
    \fontsize{9pt}{11pt}\selectfont
    \begin{tabular}{ccc|ccc}
    New Insights  & Treatment arm& Gender & 12 months & 24 months & 36 months\\\hline
    \multirow{4}{*}{Survival Rate } &  \multirow{2}{*}{FOLFOX} & Female& 0.36 (0.29,0.44) & 0.11 (0.08,0.18) &0.05 (0.02,0.09)\\
    && Male&0.32 (0.26,0.38) &0.09 (0.06,0.13)&0.04 (0.02,0.08)\\
     & \multirow{2}{*}{IROX} & Female &0.26 (0.20,0.34)& 0.05 (0.02,0.10)&0.03 (0.01,0.08)\\
     &&Male&0.25 (0.20,0.31)&0.08 (0.06,0.12) & 0.03 (0.01,0.06)\\\hline
    \multirow{4}{*}{RMST (months)}& \multirow{2}{*}{FOLFOX} & Female& 8.33 (7.81,8.99) & 10.42 (9.45,11.48)&11.33 (10.01,12.63)\\
    && Male& 8.02 (7.63,8.52) &10.37 (9.41,11.16)& 10.94 (9.91,12.12)\\
     & \multirow{2}{*}{IROX} &Female & 6.98 (6.33,7.60)&8.30 (7.32,9.45)&8.86 (7.62,10.18)\\
     &&Male&6.99 (6.44,7.47)&8.42 (7.50,9.38)& 9.05 (7.93,10.17)\\\hline
    \end{tabular}}
    \caption{Survival rates and RMST at $12,24,36$ months with $95\%$ CIs generated by SynthIPD.}
    \label{tab:new insights n9741}
\end{table}

% Additionally, the survival rates at $12$ months for each subgroup using $d_n$ and using $\tilde{d}_n$ are calculated, the results show $\mathcal{L}_{\text{sr}^{(12)}}(\tilde{d}_n\mid d_n)\leq 0.5\%$.

\begin{table}[t!]

\begin{center}
{%
    \fontsize{9pt}{11pt}\selectfont
\begin{tabular}{ccccccc}
&\multicolumn{2}{c}{FOLFOX ($n=421$)} &&\multicolumn{2}{c}{IROX ($n=383$)}& HR ($95\%$ CI)\\
 \cline{2-3}\cline{5-6}
  & Events/Total& Median PFS ($95\%$ CI)&& Events/Total& Median PFS ($95\%$ CI) & IROX vs FOLFOX\\\hline
  \textbf{By true IPD} &&&&&&\\
  \textbf{Gender} &&&&&&\\
   Male &241/247&8.2 (7.8--9.4)&&225/230&6.5 (5.8--7.8)&1.26 (1.05--1.51)\\
  Female &170/174&9.7 (8.5--11.0)&&149/153&6.9 (5.7--8.6)&1.29 (1.04--1.61)\\
  % \textbf{Age} &&&&&&\\
  % Age$<70$ & 317/324 &9.0 (8.2-10.0) & &291/300 & 6.7 (5.9-7.6) & 1.24 (1.06-1.45)\\
  % Age$\geq 70$ & 94/97 &8.2 (7.1-11.8) & &83/83& 7.1 (5.3-9.6) & 1.42 (1.06-1.92)\\
  \hline
  \textbf{By SynthIPD} &&&&&&\\
  \textbf{Gender} &&&&&&\\
   Male &241/247&8.2 (7.8--9.4)&&225/230&6.5 (5.8--7.8)&1.26 (1.05--1.51)\\
  Female &170/174&9.7 (8.5--11.0)&&149/153&6.9 (5.7--8.6)&1.29 (1.04--1.61)\\
  % \textbf{Age} &&&&&&\\
  % Age$<70$ & 311/324 &9.1 (8.2-10.1) & &286/300 & 6.7 (6.0-7.6) & 1.26 (1.07-1.47)\\
  % Age$\geq 70$ & 90/97 &8.2 (7.2-12.1) & &82/83& 7.1 (5.3-9.7) & 1.44 (1.06-1.94)\\
  \hline
\end{tabular}
}
\caption{Subgroup analysis table by Gender for N9741 study, generated by true IPD versus by synthetic IPD.}\label{tab:n9741 synthIPD}
\end{center}
\end{table}

% \begin{figure}
%     \centering
%     \includegraphics[width=0.75\linewidth]{n9741_subgroup.png}
%     \caption{The subgroup KM plots for N9741 study. The left $4$ plots are generated by true IPD, the right $4$ plots are generated by SynthIPD. Two plots in the same row correspond to the same subgroup. It's worth noting our method does not use left $4$ plots to generate IPD, they are used for \textbf{comparison purpose only}. }
%     \label{fig:n9741_subgroup}
% \end{figure}

\subsection{SynthIPD for six MM studies}\label{subsec:application 2}
We generate synthetic IPDs for all $6$ eligible studies with cytogenetic profiles (For an explicit description of the data points used, see Figure \ref{fig:6 subgroup MM} and Table \ref{tab:dara_implementation}). Then, the $6$ synthetic IPDs with cytogenetic profiles are pooled together. Then, two separate benchmarks, HR and median survival, can be established for the high-risk cytogenetic abnormalities (HRCA) subgroup with the pooled, synthetic IPD. 

The first established benchmark is the HR. The pooled HR is estimated by fitting a Cox model adjusted by cytogenetic profile and trial name. Specifically, within HRCA subgroup, the model yields pooled HRs of $0.76 (95\% \text{CI } 0.53\text{--}1.10)$ for NDMM patients and $0.43 (95\% \text{CI } 0.29\text{--}0.61)$ for RRMM patients.

\begin{remark}
    In a systematic review paper \cite{giri2020evaluation}, the use of Daratumumab is evaluated in HRCA patients to identify if such addition is associated with improved survival outcome. The same $6$ studies are included. A trial-level random effect meta-analysis is conducted in \cite{giri2020evaluation} to find the aggregated HR. Random-effects meta-analyses yield HRs of $0.76 (95\% \text{CI } 0.53\text{--}1.10)$ for NDMM and $0.43 (95\% \text{CI } 0.29\text{--}0.62)$ for RRMM. The results are comparable with our pooled analysis results, indicating the potential of utilizing SynthIPD to verify if the trial-level meta-analysis is correct.

%     \begin{table}[htbp]
%     \begin{center}
% {%
%     \fontsize{9pt}{11pt}\selectfont
% \begin{tabular}{rrrr}
% Population & Popular name & $n$(Dara vs no Dara) & HR($95\%$ CI)\\\hline
% \multirow{3}{*}{NDMM} & CASSIOPEIA \cite{moreau2019bortezomib} & 82 vs 86 &0.67 (0.35-1.30)\\
% &MAIA\cite{facon2019daratumumab} & 48 vs 44 & 0.85 (0.44-1.65)\\
% & ALYCONE \cite{mateos2018daratumumab} & 53 vs 45&  0.78 (0.43-1.43)\\\hline
% \multirow{3}{*}{RRMM}& CANDOR \cite{dimopoulos2020carfilzomib} & 48 vs 26& 0.49 (0.26-0.92)\\
% &CASTOR \cite{palumbo2016daratumumab} & 40 vs 35 & 0.41 (0.21-0.83)\\
% &POLLUX \cite{dimopoulos2016daratumumab} & 35 vs 35 & 0.34 (0.16-0.72)
% \\\hline
% \end{tabular}}
% \end{center}
%     \caption{Reported event numbers and hazard ratios (Dara versus no Dara) in high-risk cytogenetic subgroup. NDMM: newly diagnosed MM, RRMM: relapsed or refractory MM.}\label{tab:high risk cytogenetic}
% \end{table}
\end{remark}

Unlike HRs, median survival times usually have non-normal, and asymmetric sampling distributions. It is thus challenging to establish benchmark of median survival by meta-analysis. Benchmark median survivals are rarely established using meta-analysis techniques. SynthIPD offers a novel solution to overcome this limitation: the median survivals can be calculated directly by the pooled data set. The detailed median survival benchmark are established in Table \ref{tab:benchmark dara}. Utilizing the generated synthetic IPD, benchmarks can be established for different types of statistics for this specific patient population, e.g., survival rates, RMST, etc., by simple data pooling.

\begin{table}[t!]
\begin{center}
{%
    \fontsize{9pt}{11pt}\selectfont
\begin{tabular}{rrrr}
Cytogenetic profile & Disease population & Analyzed sample size & median PFS ($95\%$ CI)\\\hline
\multirow{2}{*}{HRCA} & NDMM  & 183 & NE (24.2, NE)\\
& RRMM  & 123 & 16.9 (10.2,23.7)\\
\multirow{2}{*}{Standard} & NDMM & 992 &NE (NE, NE)\\
& RRMM & 441&30.3 (26.5, 36.6)\\
\multirow{2}{*}{Unknown} & NDMM & 86&NE (NE, NE)\\
& RRMM & 285&26.1 (23.1, 38.2)
\\\hline
\end{tabular}}
\end{center}
\caption{The benchmark median PFS for patients treated by Daratumumab based on SynthIPD. NE: not estimable because the median PFS is not reached due to short follow-up.}\label{tab:benchmark dara}
\end{table}

\section{Simulation studies}\label{sec:sim}
We evaluate the performance of both DIGITIZE (only for Case (i)) and Cov-Generation via simulations. Three data-generating scenarios are considered: Case (i) the Cox proportional hazards model, Case (ii) the accelerated failure time (AFT) model, and Case (iii) the time-varying hazard model with cross-over in treatment versus control arms. Both DIGITIZE and Cov-Generation show promising accuracy, as will be illustrated below. Details of the data-generating mechanisms and
graphical illustrations are relegated to Appendix \ref{sec:simulation}. 

\begin{figure}[htbp!]
    \centering
    \includegraphics[width=\linewidth]{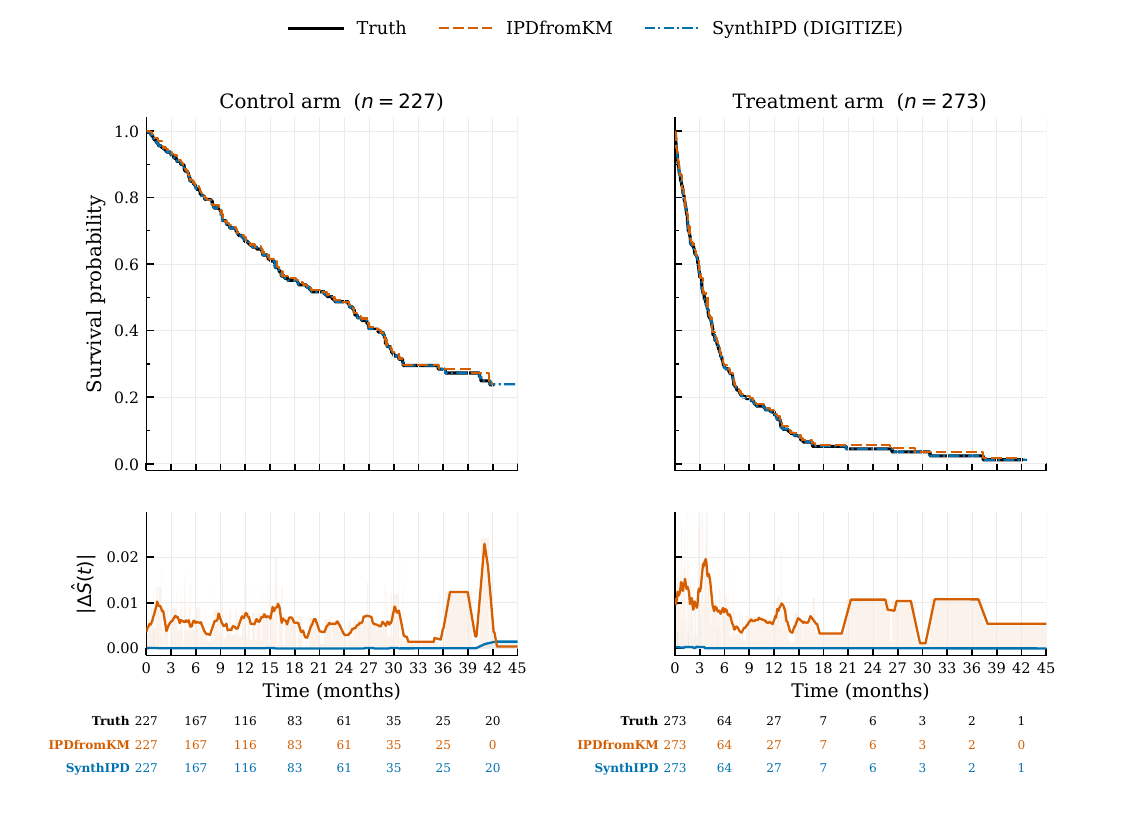}
    \caption{Comparison of SynthIPD (Blue) and IPDfromKM (Orange) method for both arms. The KM curves are visually indistinguishable from the truth for both methods. The absolute difference in survival probability over time is given in the bottom two plots. }
    \label{fig:digi comparison}
\end{figure}

For the evaluation of DIGITIZE, the IPD of Case (i) is generated once. We digitize the KM plots using both IPDfromKM \cite{liu2021ipdfromkm} and DIGITIZE. Then, we plot the KM plots for both arms side-by-side and produce the difference in survival probability $\Delta \hat{S}(t)$ in Figure \ref{fig:digi comparison}. DIGITIZE method produces synthetic data with significantly higher precision (almost $0$ error), even though both methods seem visually undistinguishable from the truth. This is expected since we can extract exact coordinates from the SVG plots, while IPDfromKM suffers from manual digitization errors. A further comparison of summary statistics generated by the three different methods is provided in Table \ref{tab:digi comparison} in Appendix \ref{subsec:digitization numerical}.

For the evaluation of Cov-Generation, SynthIPD is run $100$ times for each case. For each run, simulated IPDs of sample size $n=500$ are generated independently. Then, SynthIPDs are reconstructed for two independent covariates. We summarize the results in Table \ref{tab:results for sim}. In the table, normalized area under curve (NAUC) and Kolmogorov--Smirnov distances (KS) are used to measure the distribution similarity between simulated IPD and the corresponding synthetic IPD. Results for other metrics, such as RMST and fixed-time survival rates, are expected to be similar because these quantities are functions of the survival curve. Specifically, the normalized RMST difference is upper-bounded by NAUC, and the absolute difference in survival rate at any fixed time point is upper-bounded by KS. All three cases have negligible difference ($\lesssim 10^{-2}$) for each covariate in both distribution metrics and summary statistics. The procedure also produces stable standard deviations. For more details of the defined metrics and a visualization of the simulation results, see Appendix \ref{sec:simulation}. 

\begin{table}[htbp!]
\begin{center}{%
    \fontsize{9pt}{11pt}\selectfont
\begin{tabular}{r|rrrrr}
Case & Subgroup & $\bar{\widehat{\text{NAUC}}}(\text{SD})$ &  $\bar{\widehat{\text{KS}}}(\text{SD})$&$\bar{\mathcal{L}}_m(\tilde{d}_n)(\text{SD})$ & $\bar{\mathcal{L}}_{\tHR}(\tilde{d}_n)(\text{SD})$\\\hline
\multirow{3}{*}{CASE 1} & $X_{i,1}$ & 1.8E-2(4.2E-3) & 6.2E-2(1.7E-2) & 5.1E-3(4.7E-3) &5.2E-4(1.1E-3)  \\
&$X_{i,2}$& 1.9E-2(4.9E-3) &6.8E-2(1.7E-2) &1.1E-2(6.7E-3) & 1.0E-3(8.6E-4)\\
&$(X_{i,1},X_{i,2})$& 2.2E-2(5.4E-3) & 7.2E-2(1.5E-2) &1.7E-2(5.4E-3) & 2.0E-3(7.7E-4)  \\\hline
\multirow{3}{*}{CASE 2} & $X_{i,1}$ & 1.7E-2(5.5E-3) & 6.1E-2(1.4E-2) & 2.5E-2(9.8E-3) &5.5E-4(1.3E-3)\\
& $X_{i,2}$ & 1.9E-2(5.0E-3) & 6.2E-2(1.6E-2) &1.7E-2(3.0E-3) & 1.1E-3(1.0E-3) \\
& $(X_{i,1},X_{i,2})$ & 1.6E-2(4.7E-3) & 6.9E-2(1.2E-2)&2.1E-2(4.5E-3) & 1.8E-3(9.4E-4)  \\\hline
\multirow{3}{*}{CASE 3} & $X_{i,1}$&9.9E-3(2.3E-3) & 5.3E-2(1.1E-2) & 2.8E-2(8.1E-3) & 4.2E-3(2.7E-3)\\
 & $X_{i,2}$&1.4E-2(3.6E-3) & 5.7E-2(1.3E-2) & 1.1E-2(4.0E-3) &2.8E-3(2.4E-3) \\
 & $(X_{i,1},X_{i,2})$& 1.8E-2(6.6E-3) & 6.1E-2(1.3E-2)&1.4E-2(3.7E-3) & 3.3E-3(1.9E-3) \\
\hline
\end{tabular}}
\end{center}
\caption{The mean (SD) of $\widehat{\text{NAUC}}$, $\widehat{\text{KS}}$, $\mathcal{L}_m(\tilde{d}_n)$, and $\mathcal{L}_\tHR(\tilde{d}_n)$ for each case. Entries use scientific notation, e.g., $1\mathrm{E}\text{-}2=10^{-2}$.}\label{tab:results for sim}
\end{table}

\section{Discussion}\label{sec:conc}
This work introduces a method that reconstructs synthetic IPD only based on published KM plots and subgroup summary statistics. This work bridges the gap between access to IPD and the practical constraints that preclude data sharing. The simulation experiments and real‑world case studies demonstrate that these features translate into negligible error and downstream estimates nearly identical to those from analyses based on the original data.

SynthIPD provides a reliable alternative for true IPD, enabling downstream analyses and evidence synthesis. It also allows for accurate benchmarking of control arms across historical trials. By pooling reconstructed control-arm IPD, SynthIPD enables precise estimation of survival distributions and endpoints such as RMST. The availability of SynthIPD opens the door to methodological research areas in clinical trials and subsequent analyses.

The accuracy of SynthIPD depends on the quality and completeness of the published KM curves and summary statistics. As shown in Appendix \ref{sec:practical requirements}, missing censoring information in the risk table or missing CIs for median and/or HR may induce additional, though minor, errors. Further, the Cov-Generation method is based on SA algorithm, which is similar to greedy search. Such method is typically computationally expensive and warrants further computational improvements.

Several directions for future work remain open. First, incorporating knowledge on the correlation between covariates through Bayesian methods could extend the approach to handle higher-dimensional covariates, e.g., utilizing ideas in the population adjustment literature \cite{phillippo2020assessing}.  Second, the current Cov-Generation step is designed for discrete covariates. Extension to continuous covariates would require richer published summaries than those typically available in trial reports, for example splines, and is left for future work. Third, the generalization to generate IPD for other, non-survival, data types using similar strategy is another fruitful area of research.
% First, developing some theoretical results on convergence \cite{hajek1988cooling} and time complexity \cite{sasaki1988time} of the annealing procedure would greatly improve the method’s reliability. 
% Second, the current DIGITIZE step still requires manual intervention, automating this process would significantly enhance its practical usefulness. These areas are left for future research.}
\section*{Data availability statement}
Meta-data for the N9741 study is generated at Mayo Clinic. 
% Derived data supporting the findings of this study are available from the Dr. Qian Shi on request. 
The meta-data for the $6$ MM studies that support the findings of this study are openly available in the original publications \cite{mateos2018daratumumab,moreau2019bortezomib,facon2019daratumumab,dimopoulos2020carfilzomib,palumbo2016daratumumab, dimopoulos2016daratumumab}, at 10.1056/NEJMoa1714678; 10.1016/S0140-6736(19)31240-1; 10.1056/NEJMoa1817249; 10.1016/S0140-6736(20)30734-0; 10.1056/NEJMoa1606038; 10.1056/NEJMoa1607751 respectively.

\bibliographystyle{plain}
\bibliography{Bibliography-MM-MC}

\clearpage
\appendix
\begin{center}
  {\LARGE\bf Supplementary Materials}
\end{center}
\medskip

\section{Practical requirements of SynthIPD}\label{sec:practical requirements}

The main requirements for each step of the workflow are stated in Table \ref{tab:data-requirements}.

\begin{table}[h]
\centering
\small
\begin{tabularx}{\textwidth}{p{3.5cm} p{2.7cm} p{1.7cm} p{3cm} X}
\toprule
\textbf{Input} & \textbf{When used?} & \textbf{Required?} & \textbf{Function} & \textbf{If absent}\\
\midrule
  KM plots in SVG format for randomized patients (two arms)
  & DIGITIZE
  & Yes 
  & Extract time, survival probability, treatment arm and censorship.& Need to use IPDfromKM \cite{liu2021ipdfromkm} instead.\\\hline

  Number at-risk table
  & DIGITIZE
  & Yes
  &  Improve DIGITIZE accuracy.& Need to use IPDfromKM \cite{liu2021ipdfromkm} instead.\\\hline
  The censor table
  & DIGITIZE
  & Preferred
  & Improve DIGITIZE accuracy. & Reconstruction depend on the estimation of censoring counts.\\\midrule

  Subgroup summaries (HR and median survival)
  & Cov-Generation
  & Yes
  & Generate synthetic covariates. & The covariates cannot be generated or are not representative of the truth.\\\hline

  Confidence intervals for subgroup summaries (HR and median survival) 
  & Cov-Generation
  & Strongly Preferred
  & Generate synthetic covariates. & Reduced accuracy in Cov-Generation, the algorithm needs more time to run. \\\hline
  Subgroup summaries (Other statistics)
  & Cov-Generation
  & Preferred
  & Generate synthetic covariates more accurately. & -.\\
\bottomrule
\end{tabularx}
\caption{Practical requirements, functions and the consequences of absence.}
\label{tab:data-requirements}
\end{table}

\section{Uncertainty Quantification for DIGITIZE}\label{app:uncertainty quantification DIGITZE}

In this Section, we provide an error upper bound of the difference in survival functions between truth and synthetic IPD during the digitization step. The result separates the digitization error into $2$ distinct components: an irreducible error introduced by optimizing \eqref{eq:exact KM}, errors introduced by internal inconsistency like rounding error, low image qualities, etc.

Define the admissible space of reconstruction as
$$\mathcal{Z}:=\left\{z = (d_{\iota j},c_{\iota k}): \sum_j d_{\iota j} = D_\iota, \sum_k c_{\iota k} = C_\iota, 0\leq d_{\iota j}\leq Y_{\iota j}(z) \text{ for all }\iota\in[\ell],j\in[J_\iota]\right\}.$$
For any $z\in \mathcal{Z}$, the (estimated) KM step ratio at $s_{\iota j}$ is defined by
$$\hat{\rho}_{\iota j}(z) = 1-\frac{d_{\iota j}}{Y_{\iota j}(z)}.$$
Now, define the unobserved true KM ratios by $\rho_{\iota j}^\star$ and assume that 
$$\tilde{\rho}_{\iota j}^\star=\rho_{\iota j}^\star + \varepsilon_{\iota j},$$
is what is extracted, where $\varepsilon$ represents internal inconsistencies induced by image quality, manual extraction error, rounding, rasterization, etc. The observed, true ratio losses and censoring penalty are defined as 
\begin{align}
    &L_\rho^{\text{obs}}(z) := \sum_{\iota=1}^\ell \sum_{j=1}^{J_\iota}\left(\hat{\rho}_{\iota j}(z)-\tilde{\rho}_{\iota j}^\star\right)^2, \label{eq:observed loss}\\
     &L_\rho^\star(z) := \sum_{\iota=1}^\ell \sum_{j=1}^{J_\iota}\left(\hat{\rho}_{\iota j}(z)-\rho_{\iota j}^\star\right)^2,\label{eq:true loss}\\
     &P^\star(z) = P^{\text{obs}}(z) := \sum_{\iota=1}^\ell \sum_{k=1}^{K_\iota}\left(c_{\iota k}-\frac{C_{\iota}}{K_{\iota}}\right)^2.\label{eq:censoring penalty truth}
\end{align}
The last definition is reasonable because we assumed $C_\iota$, the number of censorings, are observed in the risk table; and $K_\iota$, the number of unique censorings, is observed without error. Then, the optimization problem in Algorithm \ref{alg:exact KM} is essentially
$$\hat{z} = \argmin_{z\in \mathcal{Z}} Q^{\text{obs}}(z),\quad Q^{\text{obs}}(z):=L_\rho^{\text{obs}}(z)+\lambda P^{\text{obs}}(z).$$
We are now ready to quantify the uncertainty in DIGITIZE step.

\begin{theorem}\label{theorem:uncertainty for DIGITIZE}
    Let $Q^\star(z) = L_\rho^\star(z)+\lambda P^\star(z)$ be the true criterion function. Then, 
    $$\sup_{t\geq 0}\left|S_{\hat{z}}(t)-S^\star(t)\right|\leq \sqrt{m \left(\inf_{z\in \mathcal{Z}}Q^\star(z)+4\sqrt{m}\|\varepsilon\|_2+2\|\varepsilon\|_2^2\right)}$$
    and for any $\tau>0$,
    $$\frac{1}{\tau}\int_0^\tau\left|S_{\hat{z}}(t)-S^\star(t)\right|dt\leq 
    \sqrt{m\left(\inf_{z\in \mathcal{Z}}Q^\star(z)+4\sqrt{m}\|\varepsilon\|_2+2\|\varepsilon\|_2^2\right)}.$$
\end{theorem}
\begin{proof}
    Note the following dervation
    \begin{align*}
        \left|Q^\star(z)-Q^{\text{obs}}(z)\right| &=\left|\sum_{\iota,j}\left(\hat{\rho}_{\iota j}(z)-\tilde{\rho}^\star_{\iota j}-\varepsilon_{\iota j}\right)^2-\sum_{\iota,j}\left(\hat{\rho}_{\iota j}(z)-\tilde{\rho}^\star_{\iota j}\right)^2\right| \\&=\left|\sum_{\iota,j}\left[2\left(\hat{\rho}_{\iota j}(z)-\tilde{\rho}^\star_{\iota j}\right)\varepsilon_{\iota j}+\varepsilon_{\iota j}^2\right]\right|
        \\&\leq 2\sum_{\iota, j}\left|\hat{\rho}_{\iota j}(z)-\tilde{\rho}^\star_{\iota j}\right||\varepsilon_{\iota j}|+\sum_{\iota, j}\varepsilon_{\iota,j}^2\\
        &\leq 2\sup_z \max_{\iota,j}\left|\hat{\rho}_{\iota j}(z)-\tilde{\rho}_{\iota j}^\star\right|\sqrt{m}\|\varepsilon\|_2+\|\varepsilon\|_2^2.
    \end{align*}
    The last inequality holds because the ratios are bounded above by $1$ and the Cauchy--Schwarz inequality (so that $\sum_{\iota,j}|\varepsilon_{\iota,j}|\leq \sqrt{m}\|\varepsilon\|_2$). By above, 
    $$Q^\star(\hat{z})\leq Q^{\text{obs}}(\hat{z})+2\sqrt{m}\|\varepsilon\|_2+\|\varepsilon\|_2^2.$$
    Using the optimality of $\hat{z}$ and the above inequality again, we have for any $z\in \mathcal{Z}$,
    \begin{align*}
        Q^\star(\hat{z})&\leq Q^{\text{obs}}(z)+2\sqrt{m}\|\varepsilon\|_2+\|\varepsilon\|_2^2\\
        &\leq Q^\star(z)+4\sqrt{m}\|\varepsilon\|_2+2\|\varepsilon\|_2^2.
    \end{align*}
    Taking infimum over $\mathcal{Z}$, 
    \begin{equation}\label{eq:bound Q^star}
        Q^\star(\hat{z})\leq \inf_{z\in \mathcal{Z}}Q^\star(z)+4\sqrt{m}\|\varepsilon\|_2+2\|\varepsilon\|_2^2.
    \end{equation} 
    For each index $r=1,\ldots, m$, let $s_r$ be the event time. We can define $\hat{\rho}_h(z) = 1-d_h/Y_h(z)$ similarly. Further, take
    $$A_r(z):=\prod_{h=1}^r\hat{\rho}_h(z),\quad B_r := \prod_{h=1}^r\rho_h^\star.$$
    By the telescoping product and taking absolute values on both sides,
    \begin{align*}
        |A_r(z) - B_r| &\leq \sum_{h=1}^r\left(\prod_{l<h}\hat{\rho}_l(z)\right)\left|\hat{\rho}_h(z)-\rho_h^\star\right|\left(\prod_{l>h}\rho_l^\star\right)\\
        &\leq \sum_{h=1}^r \left|\hat{\rho}_h(z)-\rho_h^\star\right| \leq \sum_{h=1}^m\left|\hat{\rho}_h(z)-\rho_h^\star\right|,
    \end{align*}
    where the second inequality follows by noticing $\hat{\rho}_l$ and $\rho_l^\star$ are bounded below and above by $0$ and $1$. Using Cauchy--Schwarz inequality again,
    $$\sum_{h=1}^m\left|\hat{\rho}_h(z)-\rho_h^\star\right|\leq \sqrt{m L_\rho^\star(z)}.$$
    Notice that by definition, $S_{\hat{z}}(s_r)=A_r(\hat{z})$ and $S^\star(s_r)=B_r$. Between event times, both KM curves are constant. Therefore, 
    $$\sup_{t\geq 0}\left|S_{\hat{z}}(t)-S^\star(t)\right|\leq \sqrt{m L_\rho^\star(z)}\leq \sqrt{m Q^\star(z)}.$$
    Similarly,
    $$\int_0^\tau \left|S_{\hat{z}}(t)-S^\star(t)\right|\leq \tau\sqrt{m Q^\star(z)}.$$
    The proof is finished by recalling Inequality \eqref{eq:bound Q^star}.
\end{proof}

Theorem \ref{theorem:uncertainty for DIGITIZE} provides a finite-sample deterministic uncertainty quantification for the algorithm and shows that the DIGITIZE output is stable. The error controlled by two quantities: the best achievable case inside the reconstruction class $\mathcal{Z}$ and the internal inconsistency $\varepsilon$. 

\begin{remark}
    If the published KM and risk table are internally coherent, then $\inf_z Q^\star(z)$ is small; if the extraction is nearly exact, then $\varepsilon$ is small. The utilization of SVG in our method removes most of $\varepsilon$, while other manual-based digitization methods do not. If the KM curves are coherent, then the reconstructed curve by SynthIPD must be close to the target curve, and more accurate than other manual-based digitization methods (see, e.g., Figure \ref{fig:digi comparison}). In the ideal case, we can recover the exact KM plot.
\end{remark}

\begin{remark}
    The reconstruction space assumed contains information on $C_{\iota}$ and $D_{\iota}$, indicating that the censoring numbers are reported in the risk table. In many cases, they may not be reported and only the total at risk number $C_{\iota}+D_{\iota}$ is revealed. In this case, the optimization \eqref{eq:exact KM} minimizes $L_\rho^{\text{obs}}(z)$ over a much larger set with no constraint and the upper bound will need to include a regularization bias term from $\sup_z|P^{\text{obs}}(z) - P^\star(z)|$. Since the proof is similar, we omit this part.
\end{remark}

\section{Calibration of the hyperparameters for simulated annealing algorithms}\label{app:calibration SA}
In this appendix we provide a formal derivation of the calibration rule for the simulated-annealing hyperparameters $T_{\max}$, $r^\star$, and $s$ used in the realization of $f$ given in Equation~\ref{eq:f-realization} of Section~\ref{subsec:opt}. The argument proceeds in three steps. We first establish notation and standing assumptions (Appendix \ref{app:sa-setup}). We then derive an explicit upper bound on the expected number of exploratory acceptances per outer iteration and verify its tightness (Appendix \ref{app:sa-bound}). Finally, we use this bound to motivate the recommended default values of the hyperparameters (Appendix \ref{app:calibration SA}). Throughout, we treat the analysis on the median loss $\mathcal{L}_m$ only, deferring the role of the HR loss to a remark at the end of Appendix \ref{app:calibration SA}.

\subsection{Setup and standing assumptions}
\label{app:sa-setup}

Fix an outer iteration $h \in \{1, \ldots, H\}$ of Algorithm~\ref{alg:algorithm}, and let $L_m^{(h-1)} > 0$ denote the initial median loss at the start of this iteration. Within this outer iteration, the inner loop generates up to $R$ proposals of covariates indexed by $r \in \{1, \ldots, R\}$. For each proposal $r$, write $L_m^{(h,r)}$ for the median loss of the proposed covariate vector, and define the relative loss change
\begin{equation}
    \Delta^{(h,r)} = \frac{L_m^{(h,r)} - L_m^{(h-1)}}{L_m^{(h-1)}}.
    \label{eq:app-delta}
\end{equation}
For proposals with $\Delta^{(h,r)} > 0$ (a strict worsening), the exploration branch in Algorithm~\ref{alg:algorithm} accepts proposal $r$ on the median side with probability
\begin{equation}
    \exp\!\left(-\,\frac{\Delta^{(h,r)}}{T(r)}\right),\qquad \text{where } T(r) \;=\; \frac{T_{\max}}{1 + e^{-(r - r^\star)/s}},
    \label{eq:app-acceptance}
\end{equation}
where $T_{\max}, r^\star, s > 0$ are the hyperparameters introduced in Subsection~\ref{subsec:opt} and defined in Equation~\eqref{eq:f-realization}.

Let $A^{(h,r)} \in \{0,1\}$ denote the indicator that the SA Bernoulli draw on proposal $r$ succeeds, conditional on $\Delta^{(h,r)} > 0$. The total number of proposals that explores but increases loss in iteration $h$ is given by
\begin{equation}
    N^{(h)} = \sum_{r=1}^{R} A^{(h,r)} \cdot \mathbf{1}\{\Delta^{(h,r)} > 0\}.
    \label{eq:app-Nh}
\end{equation}
The calibration of $T_{\max}$ is based on the goal of controlling the expectation $\mathbb{E}[N^{(h)}]$ under the following conditions, which may not be always true.

\begin{condition}\label{cond:worst}
    Every proposal, $r$, produces a strict worsening: $\Delta^{(h,r)} > 0$ for all $r$. This corresponds to a stuck outer iteration in which the strict-improvement branch of Algorithm~\ref{alg:algorithm} never occurs.
\end{condition}

\begin{condition}\label{cond:fixed-delta} 
    The relative loss change $\Delta^{(h,r)} \equiv \Delta$ is held fixed  for some $\Delta > 0$ across all proposals. 
    % This abstracts from the per-proposal variability of $\Delta^{(h,r)}$ to enable closed-form analysis.
\end{condition}

\begin{condition}\label{cond:hr} 
    For each outer step and each proposal, $\mathcal{L}'_{\text{HR}}<\mathcal{L}_{\text{HR}}$ in Algorithm \ref{alg:algorithm}. The derivation considers the median loss only.
\end{condition}

\begin{condition}\label{cond:hyperparam} 
    The hyperparameters satisfy $r^\star + 4s \ll R$.
    % , so that $T(r)$ reaches its asymptote $T_{\max}$ within the first few percent of inner proposals.
\end{condition}

Condition \ref{cond:worst} represents the case where the algorithm is stuck at a local minimum with respect to median RAE. Condition \ref{cond:fixed-delta} is used to simplify theoretical calibration of hyperparameters and we note that this Condition may not be true in general. Condition \ref{cond:hr} is used because in practice, RAEs for HR are relatively easy to decrease and are seldomly stuck in local minimums. Condition \ref{cond:hyperparam} is reasonable because the algorithm is similar to a greedy algorithm and $R$ is assumed to be a large number. Under Conditions \ref{cond:worst}-\ref{cond:hr}, $\{A^{(h,r)}\}_{r=1}^R$ are independent Bernoulli random variables with sucess probability$\exp(-\Delta/T(r))$. 

\subsection{Upper bound on expected acceptances}
\label{app:sa-bound}

Under Conditions \ref{cond:worst}-\ref{cond:hr}, linearity of expectation gives
\begin{equation}
    \mathbb{E}\!\left[N^{(h)}\right] \;=\; \sum_{r=1}^{R} \mathbb{E}\!\left[A^{(h,r)}\right] \;=\; \sum_{r=1}^{R} \exp\!\left(-\,\frac{\Delta}{T(r)}\right).
    \label{eq:app-exact}
\end{equation}
We establish an explicit upper bound and then verify its tightness numerically.

\begin{prop}\label{prop:bound}
Under conditions \ref{cond:worst}--\ref{cond:hr},
\begin{equation}
    \mathbb{E}\!\left[N^{(h)}\right] \;\leq\; R \cdot \exp\!\left(-\,\frac{\Delta}{T_{\max}}\right).
    \label{eq:app-bound}
\end{equation}
\end{prop}

\begin{proof}
The sigmoid $1/(1 + e^{-(r-r^\star)/s})$ takes values in $(0,1)$ for all $r \in \mathbb{R}$, so $T(r) \in (0, T_{\max})$. The function $x \mapsto \exp(-\Delta/x)$ is increasing in $x$ on $(0, \infty)$ for $\Delta > 0$. Hence
\[
    \exp\!\left(-\,\frac{\Delta}{T(r)}\right) \;\leq\; \exp\!\left(-\,\frac{\Delta}{T_{\max}}\right) \quad \text{for all } r.
\]
Summing over $r = 1, \ldots, R$ and applying~\eqref{eq:app-exact} yields the result.
\end{proof}

\begin{remark}[Tightness]
\label{rem:tightness}
Under Condition \ref{cond:hyperparam}, the bound~\eqref{eq:app-bound} is tight to some constant order. Specifically, $T(r) \geq (1-\varepsilon) T_{\max}$ whenever $r \geq r^\star + s \log\!\big((1-\varepsilon)/\varepsilon\big)$. For $\varepsilon = 0.01$, the threshold is $r^\star + 4.6 s$, which is small relative to $R$ under Condition \ref{cond:hyperparam}. Therefore, the relative difference between the bound and the expectation is
\[
    \frac{R \cdot e^{-\Delta/T_{\max}} - \mathbb{E}[N^{(h)}]}{R \cdot e^{-\Delta/T_{\max}}} \;\leq\; \frac{r^\star + 4.6 s}{R}.
\]
For the recommended defaults $T_{\max} = 0.02$, $r^\star = 3000$, $s = 2000$, and $R = 5 \times 10^5$, this gap is at most $12{,}200 / (5 \times 10^5) \approx 2.4\%$. 
% Direct numerical computation confirms this estimate: for $\Delta \in [0.005, 0.30]$, the bound overestimates~\eqref{eq:app-exact} by less than $2\%$.
\end{remark}

We motivate the recommended defaults for the simulated-annealing hyperparameters $T_{\max}$, $r^\star$, and $s$ in the realization of $f$ in Equation~\eqref{eq:f-realization}. Consider a worst-case ``stuck'' outer iteration in which every inner proposal strictly worsens the median loss by a fixed relative amount $\Delta>0$, and exploration is gated only by the median side (the HR loss is typically easy to satisfy, so we analyze $\mathcal{L}_m$ only). With $T(r)=T_{\max}/(1+e^{-(r-r^\star)/s})$, the expected number of accepted worsening moves over the $R$ inner proposals is
$$
\mathbb{E}[N] \;=\; \sum_{r=1}^{R}\exp\!\left(-\frac{\Delta}{T(r)}\right) \;\le\; R\,\exp\!\left(-\frac{\Delta}{T_{\max}}\right),
$$
since $T(r)\in(0,T_{\max})$ and $x\mapsto e^{-\Delta/x}$ is increasing on $(0,\infty)$. The bound is tight up to a relative factor of order $(r^\star+s)/R$, because $T(r)$ reaches its asymptote $T_{\max}$ within the first few thousand proposals; for the defaults below this gap is below $3\%$.

\noindent\textbf{Calibration rule.} Let $\Delta^\star>0$ be the largest relative median-loss increase the user is willing to accept at most once in expectation over a fully stuck outer iteration. Setting the upper bound equal to $1$ at $\Delta=\Delta^\star$ gives
\begin{equation}\label{eq:app-Tmax-rule}
T_{\max} \;=\; \frac{\Delta^\star}{\log R}.
\end{equation}
With $\Delta^\star=0.25$ and $R=5\times10^5$, this yields $T_{\max}=0.25/\log(5\times10^5)\approx 0.019$, which we round to $T_{\max}=0.02$; a worsening of $T_{\max}$ (resp. $3T_{\max}$) is then accepted with probability $e^{-1}\approx 0.37$ (resp. $e^{-3}\approx 0.05$). We set $r^\star=3000$ so that exploration switches on only after greedy descent stalls ($T(r)\approx 0$ while strict improvements are still found), and $s=2000$, which spreads the sigmoid activation smoothly over $r\in[1000,5000]$ (with $T(r)$ rising from $0.27\,T_{\max}$ at $r^\star-s$ to $0.73\,T_{\max}$ at $r^\star+s$). Accounting for the HR gate only multiplies the bound by a factor $q\le 1$, so the rule is conservative. These choices are robust to moderate perturbation; see Appendix~\ref{app:sensitivity}.

\section{Simulation studies and evaluations}\label{sec:simulation}
Some metrics are proposed to evaluate the DIGITIZE and Cov-Generation methods, followed by simulation verifications. Suppose both $d_n$ and $\tilde{d}_n$ are available at hand, the following two metrics are useful.
\begin{definition}[Normalized area under curves]\label{def:NAUC}
    The NAUC in subgroup $X=x,A=a$ given $d_n$ is defined as 
    $$
        \text{NAUC}_{x,a}(\tilde{d}_n\mid d_n) := \frac{1}{\tau}\int_0^\tau |\hat{S}_{x,a}(t)-\tilde{S}_{x,a}(t)|dt\in [0,1],
    $$
    where $\tau>0$ is a fixed time point, usually taken to be the largest event/censoring time.
\end{definition}

\begin{definition}[Kolmogorov--Smirnov distance]\label{def:K-S distance}
    The Kolmogorov--Smirnov distance in subgroup $X=x,A=a$ given $d_n$ is defined as
    $$\text{KS}_{x,a}(\tilde{d}_n\mid d_n):= \sup_{0\leq t\leq \tau}\left|\hat{S}_{x,a}(t)-\tilde{S}_{x,a}(t)\right|\in [0,1],$$
     where $\tau>0$ is a fixed time point, usually taken to be the largest event/censoring time.
\end{definition}

NAUC measures the averaged discrepancy between survival distributions, it is also a special case of RMST with no truncation divided by maximum follow-up time \cite{royston2013restricted}. KS distance measures the single largest vertical gap between the survival curves. For fixed $d_n$, $\tilde{d}_n$, a smaller NAUC and KS distance indicates a better synthetic copy. Both metrics can be computed using a discrete estimator. Suppose the distinct, ordered events up to time $\tau$ are $0=t_{(0)}\leq \ldots \leq t_{(q)}\leq \tau$ with $\delta_j = t_{(j)}-t_{(j-1)}, j\in[q]$, then
\begin{align*}
    &\widehat{\text{NAUC}}_{x,a}(\tilde{d}_n\mid d_n):=\frac{1}{\tau}\sum_{j=1}^{q}\delta_j\left|\hat{S}_{x,a}(t_{(j)})-\tilde{S}_{x,a}(t_{(j)})\right|,\\ &\widehat{\text{KS}}_{x,a}(\tilde{d}_n\mid d_n):=\max_{1\leq j\leq q}\left|\hat{S}_{x,a}(t_{(j)})-\tilde{S}_{x,a}(t_{(j)})\right|.
\end{align*}
Since there are $K$ categories and $2$ arms, $2K$ estimates will be produced for each metric for each synthetic data set. To see the integrated performance of $\tilde{d}_n$, it is reasonable to calculate the (weighted) average across different subgroups:
\begin{align*}
    &\widehat{\text{NAUC}}(\tilde{d}_n\mid d_n):=\sum_{x\in\{0,\ldots,K\}}\sum_{a\in\{0,1\}}\frac{n_{x,a}}{n}\widehat{\text{NAUC}}_{x,a}(\tilde{d}_n\mid d_n),\\
    &\widehat{\text{KS}}(\tilde{d}_n\mid d_n):=\frac{1}{2K}\sum_{x\in\{0,\ldots,K\}}\sum_{a\in\{0,1\}}\widehat{\text{KS}}_{x,a}(\tilde{d}_n\mid d_n).
\end{align*}

The NAUC metric is weighted by the sample size because a larger number of observations may introduce larger NAUC value. Conversely, KS distance is weighted equally because it measures the worst case scenario and each case is equally important. 

In practice, the true IPD $d_n$ is usually not available. Thus, the evaluation through NAUC or KS distance is not appropriate. One other way to evaluate the proposed method is to directly compare the summary statistics (e.g., median survivals, hazard ratios, survival rates, and their confidence intervals, etc.) generated by $\tilde{d}_n$ with that generated by $d_n$. For any summary statistics $g$, depending on its definition, the following maximum mRAEs may be convenient for this purpose
\begin{align*}
    &\mathcal{L}_g(\tilde{d}_n\mid d_n) = \max_{x,a}\left|\frac{\hat{g}_{x,a}(d_{n,x,a})-\hat{g}_{x,a}(\tilde{d}_{n,x,a})}{\hat{g}_{x,a}(d_{n,x,a})}\right|,\\
    &\mathcal{L}_{g}(\tilde{d}_n\mid d_n)  = \max_{x}\left|\frac{\hat{g}_{x}(d_{n,x})-\hat{g}_{x}(\tilde{d}_{n,x})}{\hat{g}_{x}(d_{n,x})}\right|.
\end{align*}
The two preceding losses are essentially equivalent to Equations \eqref{eq:loss1},\eqref{eq:loss2} if $g=m$ or $g=\tHR$, and can be applied to other statistics as well. Note that the calculation of $\mathcal{L}_g$ requires the summary statistics $\hat{g}_{x,a}(d_{n,x,a})$ or $\hat{g}_{x}(d_{n,x})$, to be reported in the original publication. From our experience, the survival rates $g = \text{sr}$ may be available from time to time. In the most limited case, $\mathcal{L}_m,\mathcal{L}_{\tHR}$ would be available due to our requirements in Section \ref{sec:proposed method}.

\subsection{Evaluation of DIGITIZE}\label{subsec:digitization numerical}
In this section, we provide numerical evaluation of DIGITIZE with existing digitization method IPDfromKM \cite{liu2021ipdfromkm}. Data of size $n=500$ is generated one time under CASE 1 as described in Subsection \ref{subsec:Cov-Generation numerical}. Table \ref{tab:digi comparison} shows the summary statistics given by the simulated truth and the two IPD reconstruction methods respectively. DIGITIZE method gives high accordance of summary statistics while IPDfromKM may suffer from manual extraction error (e.g., in Table \ref{tab:digi comparison}, there is an $11\%$ mRAE in median PFS for the control arm). In our numerical studies, we generally observe from $5\%$ to up to $20\%$ maximum mRAEs on average for IPDfromKM while the mRAEs are below $1\%$ for DIGITIZE.

\begin{table}[t!]
\begin{center}
{%
    \fontsize{9pt}{11pt}\selectfont
\begin{tabular}{r|rrrr}
Method & Treatment & events/n & median PFS ($95\%$ CI) & HR($95\%$ CI)\\\hline
\multirow{2}{*}{Truth} & Treatment &121/227 &22.55 (16.37--26.93) & 0.21 (0.16--0.26)\\
 & Control& 224/273 & 3.55 (3.08--4.34) & Reference level\\\hline
 \multirow{2}{*}{IPDfromKM}& Treatment& 119/227  &22.83 (16.52--27.02) & 0.21 (0.16--0.27) \\
 & Control & 220/273& 3.95 (3.09--4.43) & Reference level\\\hline
 \multirow{2}{*}{SynthIPD} & Treatment& 121/227  &22.55 (16.37--26.93) & 0.21 (0.16--0.27)\\
 & Control &224/273& 3.55 (3.08--4.34) &Reference level\\
 \hline
\end{tabular}}
\end{center}
\caption{The comparison of summary statistics with different methods for digitization.}\label{tab:digi comparison}
\end{table}

\subsection{Evaluation of Cov-Generation}\label{subsec:Cov-Generation numerical}
To evaluate the performance of Cov-Generation, we simulate IPD under multiple data-generating mechanisms and assume a `perfect' digitization process. Specifically, we set \(\tilde{u}_i = u_i\), \(\tilde{\delta}_i = \delta_i\), and \(\tilde{a}_i = a_i\) for every \(i \in \{1, \ldots, n\}\), and generate only the synthetic covariates \(\tilde{x}_i\) such that \(\tilde{d}_n = \{u_i, \delta_i, a_i, \tilde{x}_i\}_{i=1}^{n}\). Because the true IPD \(d_n\) is known from simulation, the performance of the method can be directly evaluated using all proposed metrics. For the results reported in Section \ref{sec:sim}, Appendices \ref{subsec:Cov-Generation numerical} and \ref{app:sensitivity}, the simulations are based on the following steps.

\begin{enumerate}
    \item Run the Cov-Generation algorithm $B=100$ times, using the same $\tilde{d}_n^{(-x)}$.
    \item Retain all $B$ covariate configurations $x^{(1)},\ldots, x^{(B)}$.
    \item Store the data $\tilde{d}_n^{(b)}$.
    \item Compute the metrics $\text{NAUC}^{(b)}, \text{KS}^{(b)}, \mathcal{L}_m^{(b)}, \mathcal{L}_{\text{HR}}^{(b)}, \mathcal{L}_{\text{sr}3}^{(b)}$ for each repetition $b=1,\ldots, B$. Aggregate the summaries (mean and SD) over $B$ repetitions.
\end{enumerate}

Three distinct data-generating mechanisms are considered. We apply the procedure \(\text{Cov-Generation}(\tilde{d}_{500}^{(-x)}, \hat{\mathcal{J}}_{\hat{m}_{x,a}}, \hat{\mathcal{J}}_{\widehat{\tHR}_x}; 50, 5\times10^5, \alpha, \beta, f, 0.02, 0)\), where $f\equiv 0$ or $f=f_{sa}$ as defined in Equation \eqref{eq:f-realization}. Setting \(S_m = 0\) disables early termination based on loss thresholds, ensuring that each simulation terminates only after completing 50 loss updates.

\noindent\textbf{CASE (i). (Proportional hazard model)} Consider two covariates $X_{i,1},X_{i,2}$ both simulated independently and identically from $\text{Bernoulli}(0.3)$ distributions. The responses are generated from a Cox proportional hazards model $$\lambda(t\mid A_i,\vX_i) = \frac{\log(2)}{10}\exp\left(\log(0.7)A_i+\log(0.5)X_{i,1}+\log(0.75)X_{i,2}\right).$$

\noindent \textbf{CASE (ii). (AFT model)} Let the two covariates $X_{i,1},X_{i,2}$ be independent random variables from Bernoulli$(0.3)$, Bernoulli$(0.4)$ respectively. The event time follows an accelerated failure time (AFT) model $\log(T_i) = 0.8 A_i -0.7 X_{i,1}+0.5 X_{i,2}+\epsilon_i$
where $\epsilon_i\sim \mathcal{N}(0,0.3^2)$. 

\noindent \textbf{CASE (iii). (Cross-over)} Consider two covariates $X_{i,1},X_{i,2}$ both simulated independently and identically from $\text{Bernoulli}(0.3)$ distributions. Consider a case under which the baseline hazards vary by time. Specifically, set $t_0 = 5,$ the baseline hazard for group $a$ is defined as
    $$
      \lambda_a(t)=0.1\mathbbm{1}\{a=0,\;t<t_0\}
+0.2\mathbbm{1}\{a=1,\;t<t_0\}
+0.3\mathbbm{1}\{a=0,\;t\ge t_0\}
+0.1\mathbbm{1}\{a=1,\;t\ge t_0\}$$
where $\mathbbm{1}(\cdot)$ is the indicator function. The individual hazard function is then defined as
$\lambda(t\mid A_i=a,\vX_i)= \lambda_a(t)\,\exp(\log(1.8)X_{i,1} + \log (0.5)X_{i,2})$.

\begin{figure}[htbp]
    \centering
    \includegraphics[width=1\linewidth]{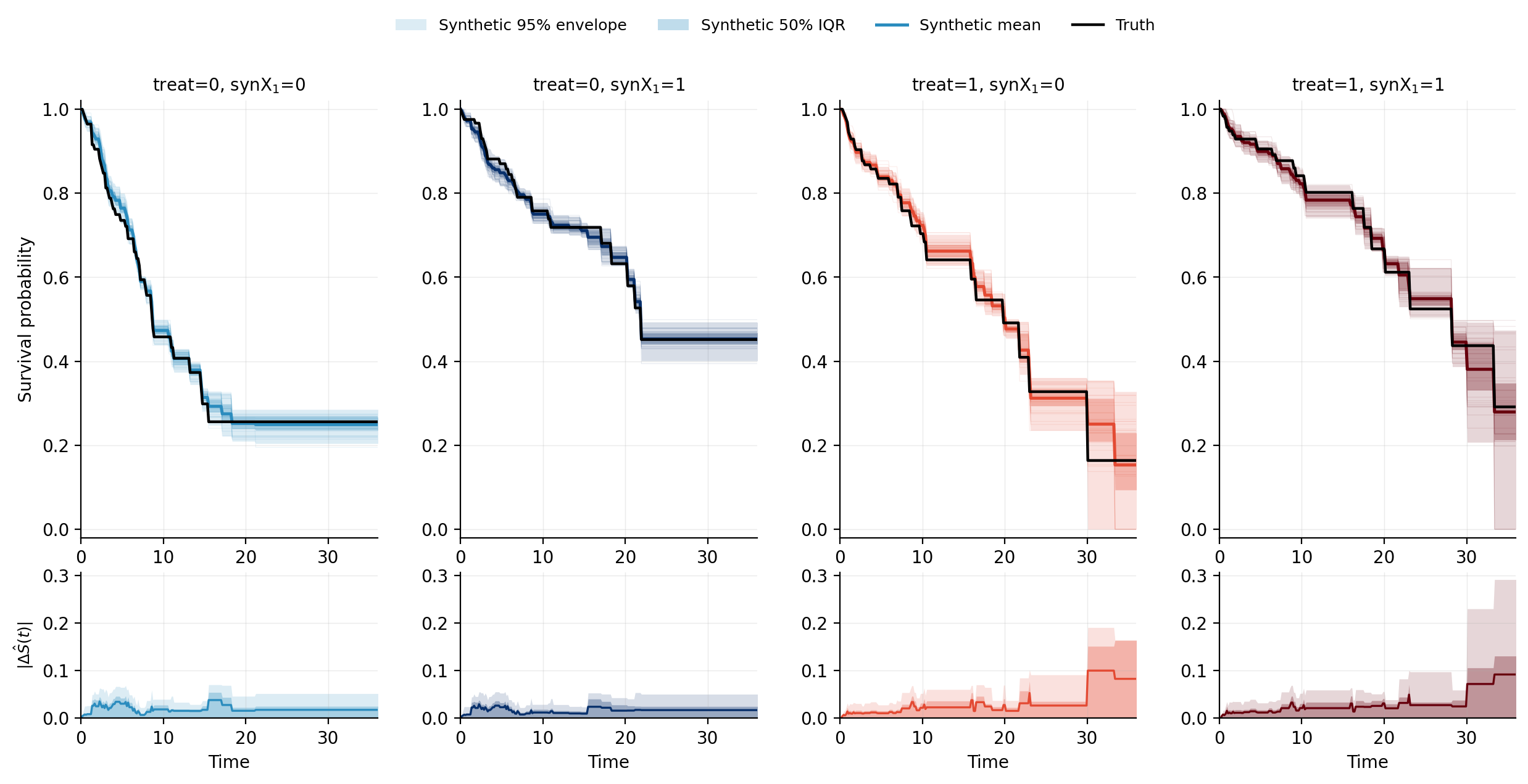}
    \caption{A figure illustration of the uncertainty of Cov-Generation method for CASE 1 with $\alpha=0.05,\beta=0.2, f\equiv 0$. \textbf{Solid black lines}: Simulated truth; \textbf{Solid colored lines}: Mean of $\hat{S}_r(t_j\mid x_1,a)$ over $r=1,\ldots,100$ repetitions for each fixed $t_j$; \textbf{Dark shaded areas}: $25\%-75\%$ percentile band for $\hat{S}_r(t_j\mid x_1,a)$; \textbf{Light shaded areas}: $95\%$ percentile band for $\hat{S}_r(t_j\mid x_1,a)$.}
    \label{fig:km overlay}
\end{figure}

CASE 1 examines the popular Cox model, CASE 2 studies the accelerated failure time (AFT) model, under which the Cox assumption is violated. CASE 3 investigates a complex scenario where there is a cross-over in survival time, i.e., the KM plots for the two treatment groups intersect at one or more time points. For each case, the censoring times will be generated by an exponential distribution such that the censoring rate is between $30\%-50\%$. Additionally, we visualize the KM uncertainties of our method for CASE 1 in Figures \ref{fig:km overlay} and its discussions.

In Table \ref{tab:results for sim}, we observe that the standard deviation of \(\bar{\mathcal{L}}_{\tHR}(\tilde{d}_n)\) is relatively large compared to its mean (\(10^{-4}\) vs. \(10^{-3}\)). 
This variability primarily arises because the total number of loss updates \(H\) limits the optimization, causing some simulations to terminate before the \(\mathcal{L}_{\tHR}\) value has fully converged. 
For instance, among the 100 repetitions of \(X_{i,1}\) in Case~1, more than 90 runs achieve \(\mathcal{L}_{\tHR}\) at level of \(10^{-4}\) or lower, while a small proportion still yield losses in the range of \(10^{-3}\) to \(10^{-2}\) which will increase the standard deviation. 
Increasing \(H\) would further reduce this variability by allowing additional optimization updates and more complete convergence. 
Nonetheless, even with the current setting, an HR loss below 0.01 already represents a sufficiently accurate reconstruction for practical purposes.

\begin{figure}[htbp]
    \centering
    \includegraphics[width=1\linewidth]{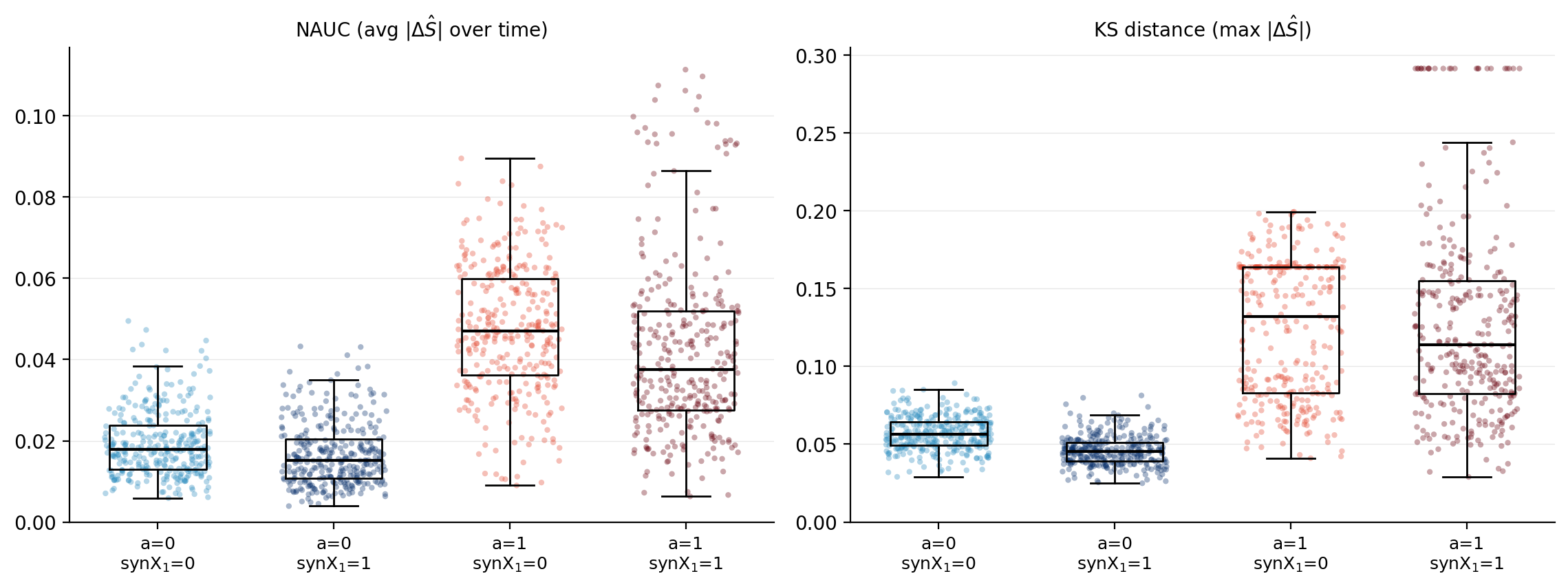}
    \caption{The scatter plot of the NAUC and KS distance metrics over the $100$ runs for CASE 1.}
    \label{fig:scatter box plot}
\end{figure}

To visualize the uncertainty of Cov-Generation over $100$ repetitions, we create a fixed grid of $300$ equally-spaced time points from $0$ to the maximum observed time. Then, a survival probability matrix of dimension $100\times 300$ with rows $\hat{S}_1(t_j|x_1,a),\ldots, \hat{S}_{100}(t_j|x_1,a)$ is constructed, where $\hat{S}_r(t_j|x_1,a)$ is the survival probability at time $t_j$ of the $r$th repetition for covariate $x_1$ and arm $a$. For each time point, we extract the $2.5\%,25\%,75\%,97.5\%$ percentile of the survival function together with the mean. The results are summarized and reported in Figures \ref{fig:km overlay}, \ref{fig:scatter box plot}. 

The reported lines in Figure \ref{fig:km overlay} can be viewed as a sufficiently smoothed approximation of the KM curves, averaged over $100$ repetitions. One can observe that the method is stable across multiple runs and $|\Delta \hat{S}(t)|$ is small for most of the time points except at the tails. In Figure \ref{fig:scatter box plot}, the NAUC and KS distance for each run under CASE 1 using $f\equiv 0$ are reported as distinct data points. Most results lie in a good range, indicating very low difference between SynthIPD and the truth. There are several cases where the KS distance is $30\%$ off for $X_1=1,a=1$. This is due to the estimation error at the tail of KM curves, which will make large impact to the KS statistic even there is only a limited number of discrepancies.


\subsection{Sensitivity analysis for hyperparameters}\label{app:sensitivity}
In this section, we perform several sensitivity analyses to assess the effect of the
hyperparameters on the algorithm's performance. With a single covariate and only a few
levels, the algorithm always converges rapidly and reaches the termination threshold; we therefore employ a dataset with more levels and a larger sample size. In each run, an i.i.d IPD of sample size $n=1200$ with $3$ covariates -- ECOG PS of $0$ or $1$, Disease Stage of Metastatic or Advanced, PD-L1 Expression Level of High or Low --
forming $8$ strata is generated via a complex time-varying Cox model with treatment-strata interactions:

The survival time $T_i$ follows hazard function $\lambda(t\mid A_i,\vX_i)=\lambda_0\alpha t^{\alpha-1}\exp(\bm{\beta}^\top \bm{Z}_i)$ where $\bm{Z}_i=(A_i,X_{i,1},X_{i,2},X_{i,3},A_iX_{i,3})^\top$ is the design vector, $\bm{\beta}=(-0.4,0.5,0.45,-0.2,-0.25)^\top$, $\alpha=1.5,\lambda_0=0.01$. The censoring time $C_i$ follows Exponential distribution with mean $0.012$. Additionally, the subjects will be censored administratively after $36$ months, yielding a final censoring rate $\approx 73\%$. The assignments $A_i$ are generated by stratified permuted block design for each stratum.

We first investigate the impact of the choice of
$(\alpha,\beta)$ by considering the following cases:
\[
(\alpha,\beta) = (1\%,5\%), (5\%,20\%),(10\%,40\%),(20\%,50\%).
\]
We then fix $(\alpha,\beta)=(5\%,20\%)$ and compare the performance of the SA and
non-SA algorithms in terms of convergence time. Here $f$ is set to the recommended
values in Subsection~\ref{subsec:opt}.

\subsubsection{Sensitivity analysis for 
$(\alpha,\beta)$}
\label{appdx:sub_section_ab}
In this subsection, we set $H = 50$ and $R = 2\times 10^5$, and record the errors as a function of cumulative proposals, $r$. Each scenario is repeated for $96$ times. Table~\ref{tab:ab-sensitivity} summarizes the sensitivity of Cov-Generation to the
perturbation rate $(\alpha,\beta)$ on the dataset. At every cumulative-proposal
budget, the loss decreases monotonically as $(\alpha,\beta)$ becomes smaller. This is anticipated by Section~\ref{subsubsec:cov-gen}: a large $\beta-\alpha$ perturbs heavily for each proposal, producing moves that are
rejected with a higher probability, which slows the descent time. 

\begin{table}[t!]
\centering
\caption{Sensitivity of Cov-Generation to the perturbation rate $(\alpha,\beta)$ on the dataset ($8$ strata, $n=1200$). Entries are the mean (SD) across $96$ independent replications of the median loss $\mathcal{L}_m$ and the hazard-ratio loss
$\mathcal{L}_{\mathrm{HR}}$ (both mRAEs), evaluated at fixed cumulative-proposal budgets.
Bold marks the lowest mean error in each column.}
\label{tab:ab-sensitivity}
\begin{tabular}{l ccc}
\toprule
& \multicolumn{3}{c}{Cumulative proposals} \\
\cmidrule(lr){2-4}
$(\alpha,\beta)$ & $10^{3}$ & $10^{5}$ & $10^{6}$ \\
\midrule
\multicolumn{4}{l}{\emph{Median loss} $\mathcal{L}_m$}\\
$(1,5)$  & \textbf{0.252 (0.040)} & \textbf{0.173 (0.031)} & \textbf{0.150 (0.034)} \\
$(5,20)$  & 0.273 (0.037) & 0.209 (0.032) & 0.186 (0.031) \\
$(10,40)$ & 0.285 (0.043) & 0.225 (0.034) & 0.206 (0.031) \\
$(20,50)$ & 0.288 (0.044) & 0.232 (0.036) & 0.213 (0.032) \\
\addlinespace
\multicolumn{4}{l}{\emph{Hazard-ratio loss} $\mathcal{L}_{\mathrm{HR}}$}\\
$(1,5)$  & 0.197 (0.061) & \textbf{0.091 (0.038)} & \textbf{0.068 (0.025)} \\
$(5,20)$  & \textbf{0.181 (0.054)} & 0.094 (0.034) & 0.072 (0.022) \\
$(10,40)$ & 0.203 (0.053) & 0.118 (0.036) & 0.092 (0.029) \\
$(20,50)$ & 0.218 (0.061) & 0.144 (0.054) & 0.119 (0.047) \\
\bottomrule
\end{tabular}
\end{table}

That said, we do not commit to a single $(\alpha,\beta)$ in practice. Since SynthIPD is training-free
and each reconstruction is inexpensive and independent, we run several $(\alpha,\beta)$ configurations in parallel and retain the reconstruction with the smallest RAEs
$(\mathcal{L}_m,\mathcal{L}_{\mathrm{HR}})$. The sensitivity reported above thus
informs, but does not constrain, the final output.

\subsubsection{Sensitivity analysis on optimization algorithm}
\label{appdx:sub_section_sa}
To evaluate the convergence of the algorithm, SynthIPD is run for $96$ times under each of the SA and non-SA settings. Here, $f$ is set to the recommended setup in Subsection~\ref{subsec:opt}. The hyperparameters are set as follows: $\alpha = 1\%,\beta = 5\%, H = 100, R = 2\times 10^5, S_m= 10^{-8}, S_{\text{HR}} = 0.2$. Note that $S_m$ is set to a near-zero value to disable early termination due to small median loss, while $S_{HR}$ is set to a loose threshold. Under this
configuration, the acceptance criterion in Algorithm~\ref{alg:algorithm} reduces
to requiring only a strict improvement in $\mathcal{L}_{HR}$ at each update,
so that the optimizer focuses on driving down $\mathcal{L}_m$.

\begin{figure}[t!]
  \centering
  \includegraphics[width=\linewidth]{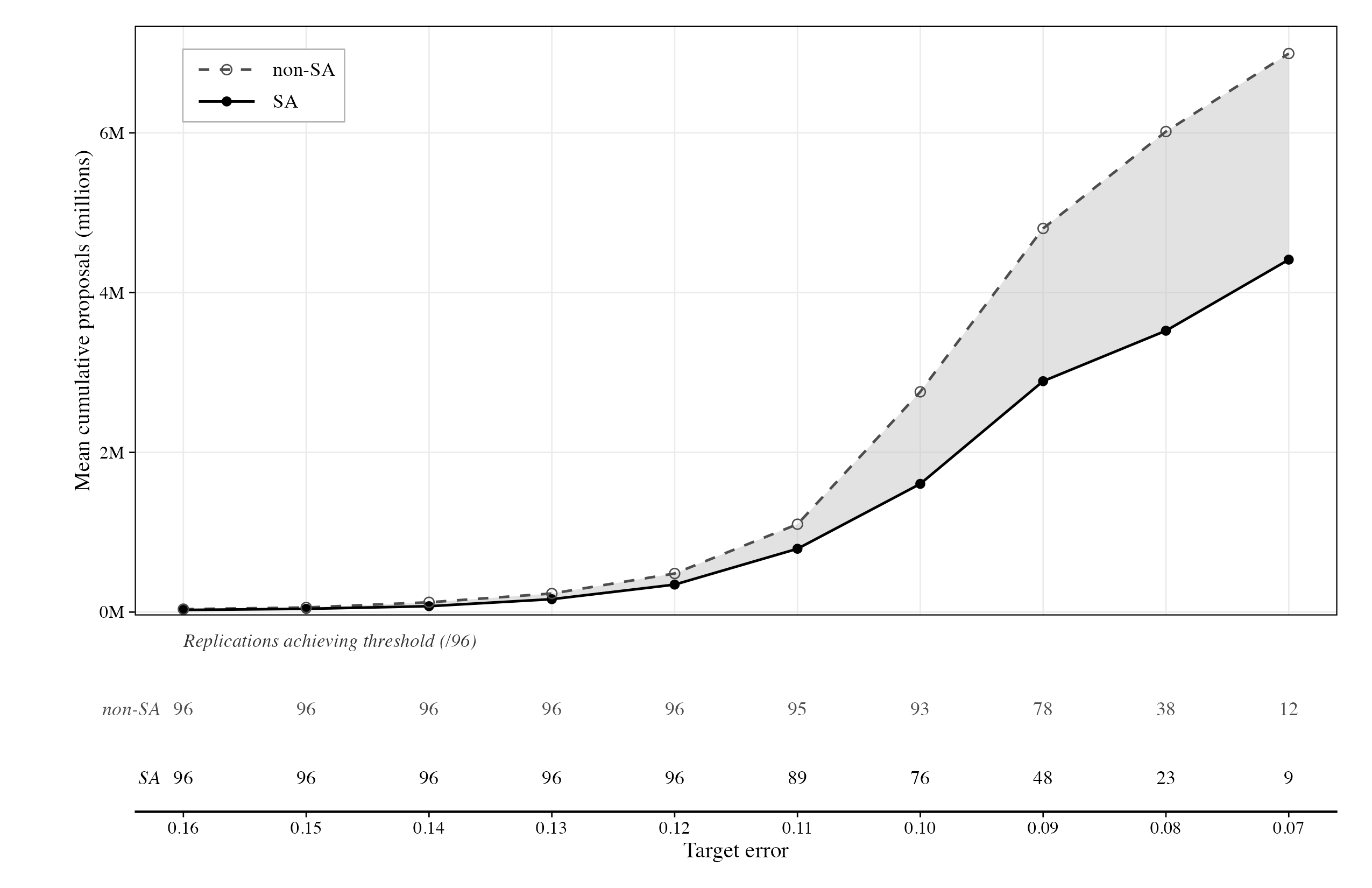}
  \caption{Comparison of SA and non-SA variants of Cov-Generation across 96
  independent replications.
  \textbf{Top curves:} mean cumulative proposals required to first reach
  target error between $[0.07,0.16]$, averaged over replications that reached the target within the iteration budget; the shaded region indicates the number of proposals (iterations) saved by SA on average. \textbf{Bottom table:} number of replications
  (out of 96) that cannot reach the target error. }
  \label{fig:sa-vs-nonsa}
\end{figure}

The results of the two settings are reported in Figure~\ref{fig:sa-vs-nonsa}. Among the 96 replications under each setting, no run achieved an error below $0.06$. On the speed of convergence, the two settings show a substantial discrepancy: SA significantly reduces the mean number of proposals required to first reach each threshold. In practice, the algorithm is typically stopped once a pre-specified threshold is reached, so consumed time is approximately proportional to the cumulative-proposal counts reported in the $y$-axis of Figure \ref{fig:sa-vs-nonsa}.

\subsection{SynthIPD as pseudo IPD}\label{app:pseudo IPD}

In this section, we validate the performance of our synthetic data as training data for
several synthetic-data generative models. We compare three scenarios as follows:
 
\begin{align}
    &\text{IPD} \overset{\text{Generative modeling}}\to \text{Synthetic data},\label{eq:conventional generative}\\
    &\text{IPD} + \text{SynthIPD} \overset{\text{Generative modeling}}\to \text{Synthetic data},\label{eq:augmentation}\\
    &\text{SynthIPD} \overset{\text{Generative modeling}}\to \text{Synthetic data}.\label{eq:privacy}
\end{align}
Here, pipeline \eqref{eq:conventional generative} is the conventional training process of generative models. The second \eqref{eq:augmentation} generates SynthIPD from IPD and the data is combined for training, we view this as an application for data augmentation. This is especially suitable when the generative model requires a larger piece of IPD. The last one \eqref{eq:privacy} depicts the case where strict privacy is needed for original IPD, so only summary level information like KM plots and summary statistics can be shared. We note that in this application, there is no longer restriction on the reporting format of summary statistics because, in principle, any level of within-strata summary is transferable without leaking privacy.

\begin{table}[t!]
\centering
\small
\begin{tabular}{cccc}
\toprule
\textbf{Name} & \textbf{Method} & \textbf{Citation} & \textbf{Code source} \\
\midrule
   SynthIPD & Optimization & - &-\\
   survivalGAN & GAN & \cite{norcliffe2023survivalgan} &Python Package ``synthcity''\\
   CTGAN & GAN & \cite{xu2019modeling} &Python Package ``sdv''\\
   TVAE & VAE & \cite{xu2019modeling} &Python Package ``synthcity''\\
   SurvDiff & Diffusion & \cite{brockschmidt2026survdiffdiffusionmodelgenerating} &GitHub\\
   Conditional tree & CART & \cite{azizi2021can} &R Package ``synthpop'' \\
\bottomrule
\end{tabular}
\caption{A summary of generative models used in our investigation. GAN: generative adversarial network; CTGAN: conditional tabular GAN; VAE: variational auto-encoder; TVAE: tabular VAE; CART: classification and regression trees.}
\label{tab:generative models}
\end{table}

\begin{figure}[p]
  \centering
  \includegraphics[width=\textwidth]{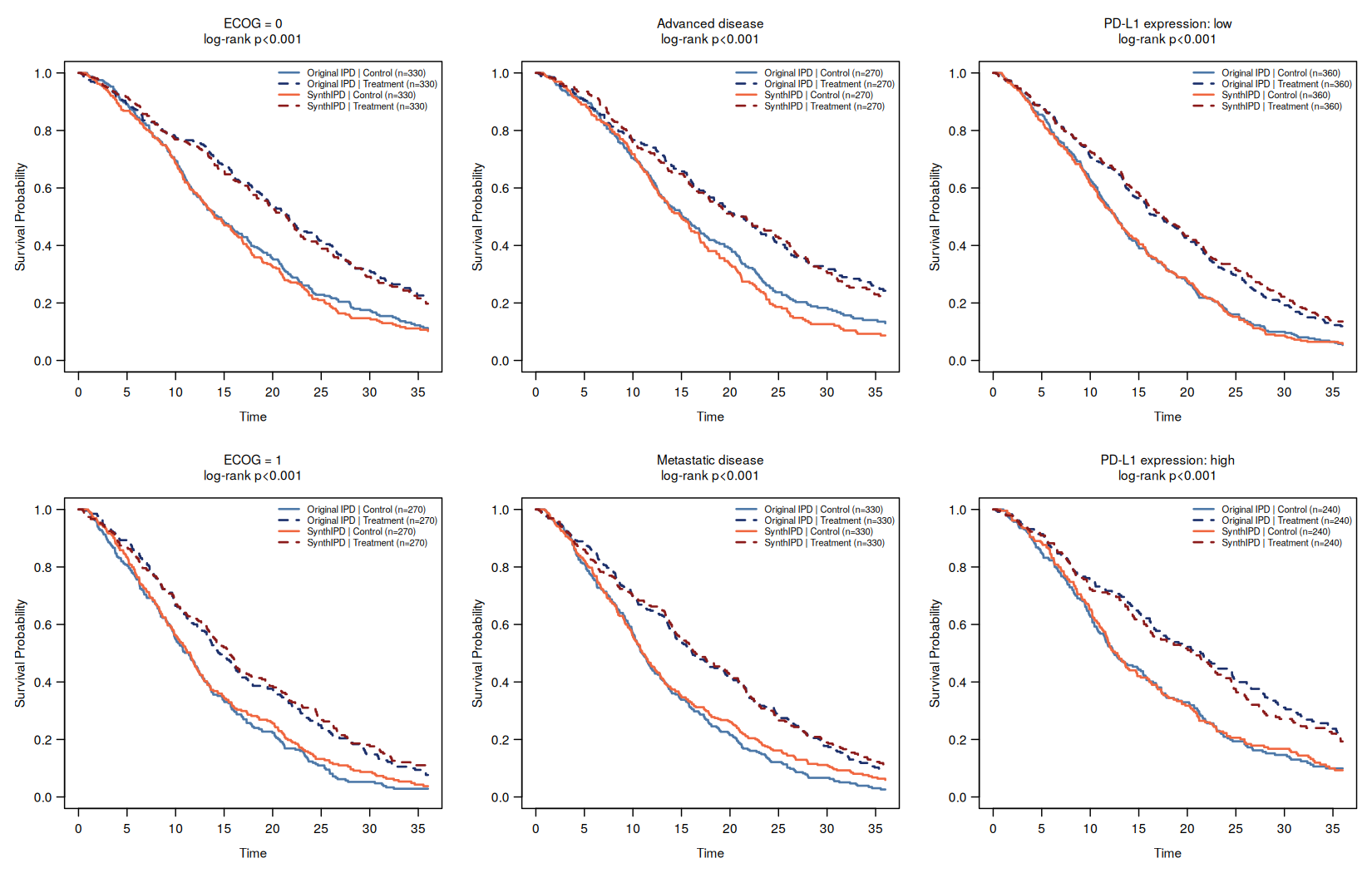}
  \caption{Kaplan--Meier curves comparing the original IPD and the selected SynthIPD
  realization, separately by treatment arm, within marginal subgroups defined by ECOG
  performance status (left column), disease stage (advanced/metastatic, middle column),
  and PD-L1 expression level (low/high, right column). }
  \label{fig:km_marginal}
\end{figure}

To set up the comparison across the three scenarios, we use the same original IPD
($n = 1200$, three covariates) as in Subsection~\ref{app:sensitivity}. We select a single
SynthIPD realization from the replications of the previous simulation, which attains
mRAEs of $\mathcal{L}_m = 0.06$ and $\mathcal{L}_{\mathrm{HR}} = 0.11$. Figure~\ref{fig:km_marginal} reports the KM comparison between the original IPD
and this SynthIPD realization at the marginal levels of ECOG performance status, disease
stage, and PD-L1 expression level by treatment arms. 

We consider five different generative models, listed in
Table~\ref{tab:generative models}. To further evaluate whether SynthIPD can serve as a reliable training set for generative
modeling, we compare the discriminative performance of synthetic datasets generated under
the three scenarios. The C-index is used because it measures whether the generated data preserve the relative ordering of patients' survival risks implied by the baseline covariates. The discrepancies are quantified as 
\begin{align*}
\Delta
&=
\left| C_{\text{Synthetic method}} - C_{\mathrm{ref}} \right|,
\end{align*}
where the baseline C-index is computed from the original dataset and equals
$C_{\text{ref}}= 0.6219$. A small $\Delta$ indicates that the synthetic data preserve the properties of the original IPD. Thus, comparing the C-index discrepancies across these
three training scenarios provides evidence on whether SynthIPD is sufficiently faithful to the original IPD for downstream synthetic data generation. The discrepancies $\Delta$ are reported in Table~\ref{tab:cindex_discrepancy_models}. Among all five generative models, the combined
training (IPD + SynthIPD) achieves the best performance for four of them.

\begin{table}[htbp]
\centering
\caption{Comparison of C-index discrepancies across generative models. }
\label{tab:cindex_discrepancy_models}
\small
\begin{tabular}{llcccccc}
\toprule
\multirow{2}{*}{Model} & \multirow{2}{*}{Class}
& \multicolumn{2}{c}{Original IPD}
& \multicolumn{2}{c}{SynthIPD}
& \multicolumn{2}{c}{IPD + SynthIPD} \\
\cmidrule(lr){3-4} \cmidrule(lr){5-6} \cmidrule(lr){7-8}
& & C-index & $\Delta$ & C-index & $\Delta$ & C-index & $\Delta$ \\
\midrule
survivalGAN      & GAN       & 0.6192 & 0.0027 & 0.6101 & 0.0118 & 0.6224 & \textbf{0.0005} \\
CTGAN            & GAN       & 0.6136 & 0.0083 & 0.6178 & 0.0041 & 0.6179 & \textbf{0.0040} \\
TVAE             & VAE       & 0.6155 & 0.0064 & 0.6159 & 0.0060 & 0.6213 & \textbf{0.0006} \\
SurvDiff         & Diffusion & 0.6153 & 0.0066 & 0.6131 & 0.0088 & 0.6174 & \textbf{0.0045 }\\
Conditional tree & CART      & 0.6104 & \textbf{0.0115} & 0.6078 & 0.0141 & 0.5985 & 0.0234 \\
\bottomrule
\end{tabular}
\end{table}

The result for conditional tree is different from conventional neural network based generative models. The conditional tree is a low-variance conditional resampler that reproduces its training distribution while keeping the bias. The neural network-based models instead estimate a smoothed representation, so at the present
sample size ($n=1200$) they are variance-limited and comparatively robust to noise in the
training data. For these models
the additional samples reduce estimation variance, so the combined set improves on either source alone. For the conditional tree the variance reduction is limited. Thus, mixing in SynthIPD only transmits its slightly attenuated
individual-level association into the output, and the original IPD remains best.

\begin{figure}[t!] \centering \includegraphics[width=\textwidth]{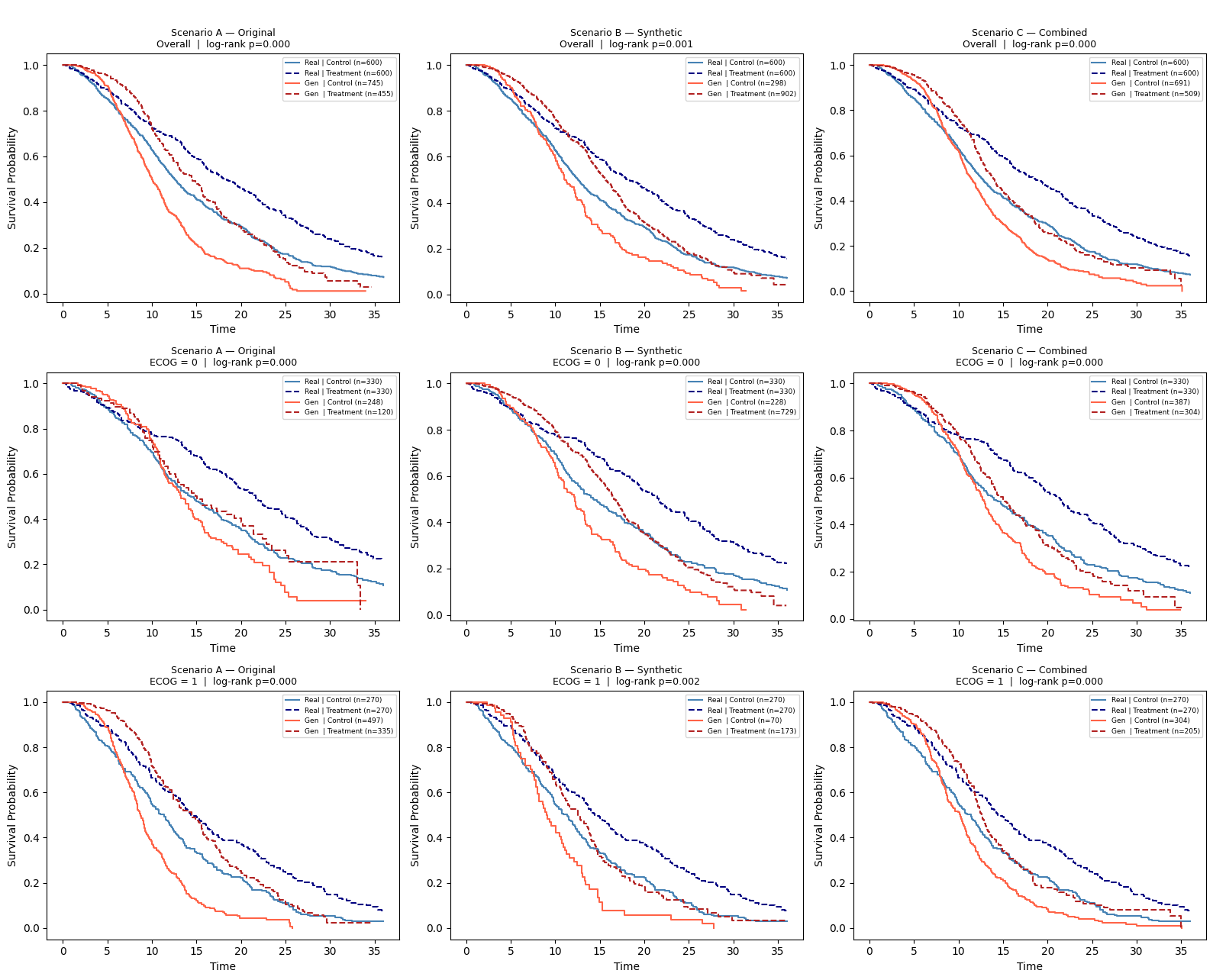} \caption{KM diagnostic plots comparing the real IPD and SurvivalGAN-generated datasets under three training scenarios: original IPD, SynthIPD, and IPD + SynthIPD combined. ECOG is the hypothetical subgroup name taking values in $\{0,1\}$.} \label{fig:km_survivalgan} \end{figure}

\begin{figure}[t!] \centering \includegraphics[width=\textwidth]{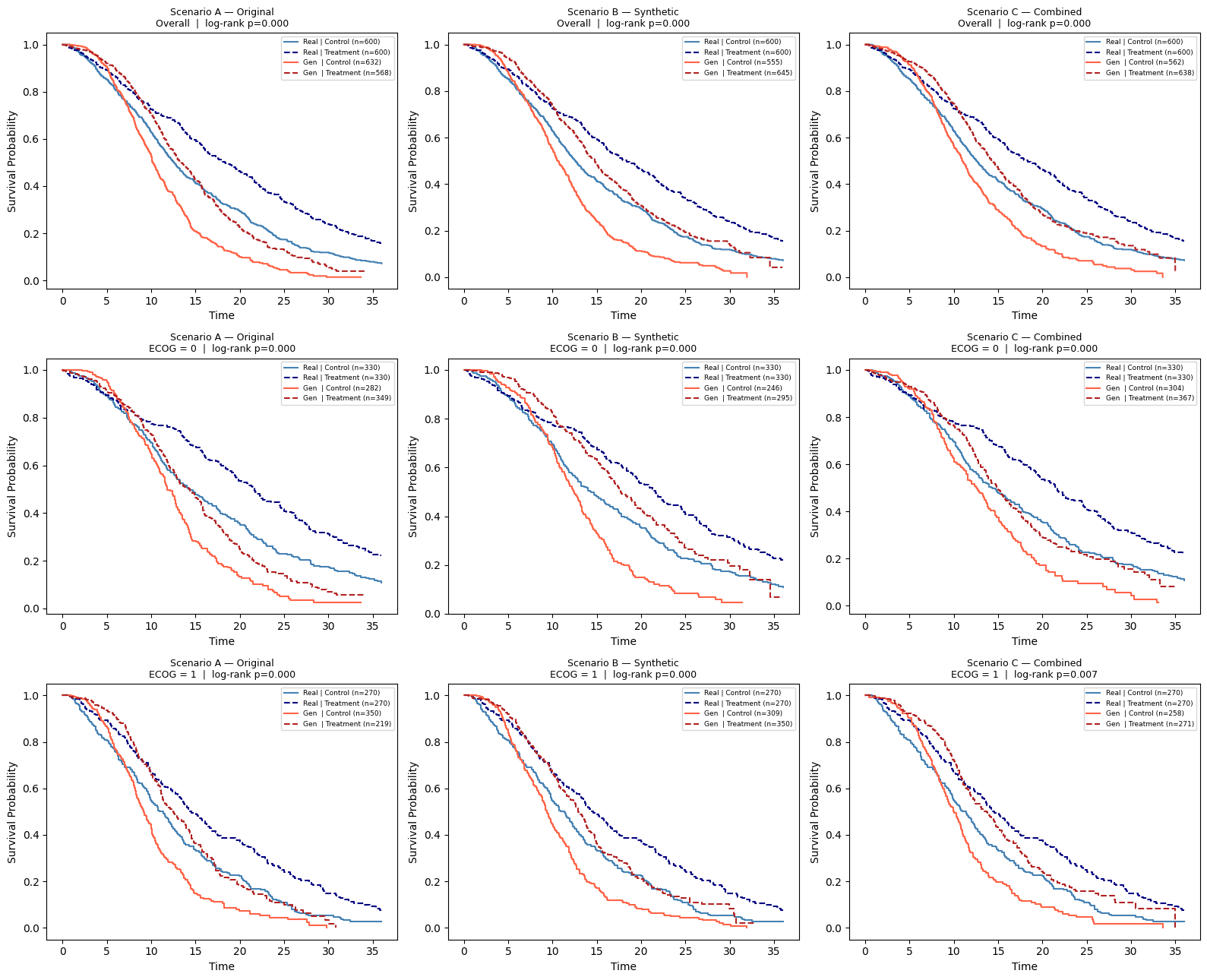} \caption{KM diagnostic plots comparing the real IPD and CTGAN-generated datasets under three training scenarios: original IPD, SynthIPD, and IPD + SynthIPD combined. ECOG is the hypothetical subgroup name taking values in $\{0,1\}$.} \label{fig:km_ctgan} \end{figure}

\begin{figure}[t!] \centering \includegraphics[width=\textwidth]{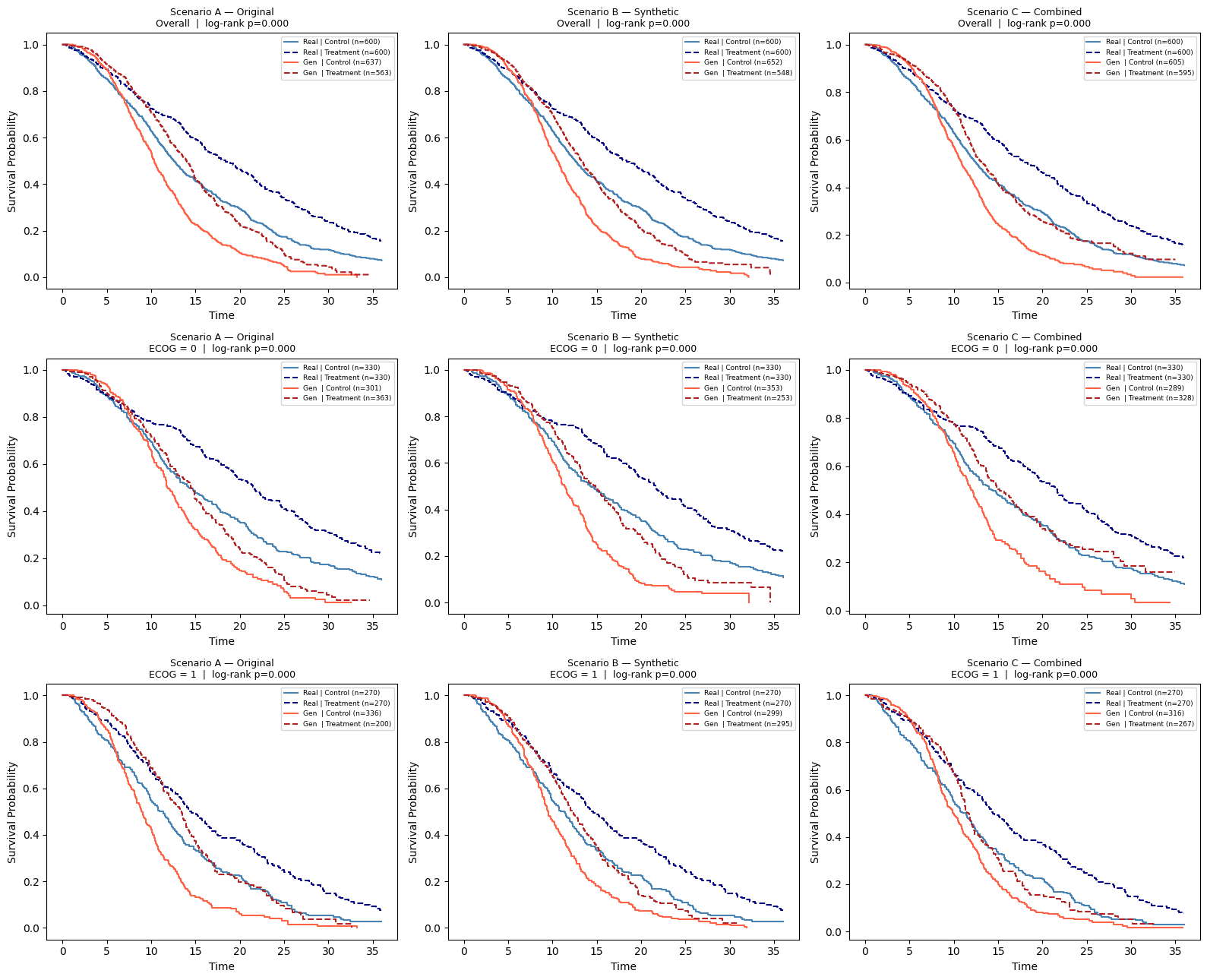} \caption{KM diagnostic plots comparing the real IPD and TVAE-generated datasets under three training scenarios: original IPD, SynthIPD, and IPD + SynthIPD combined. ECOG is the hypothetical subgroup name taking values in $\{0,1\}$.} \label{fig:km_tvae} \end{figure}

\begin{figure}[p] \centering \includegraphics[width=\textwidth]{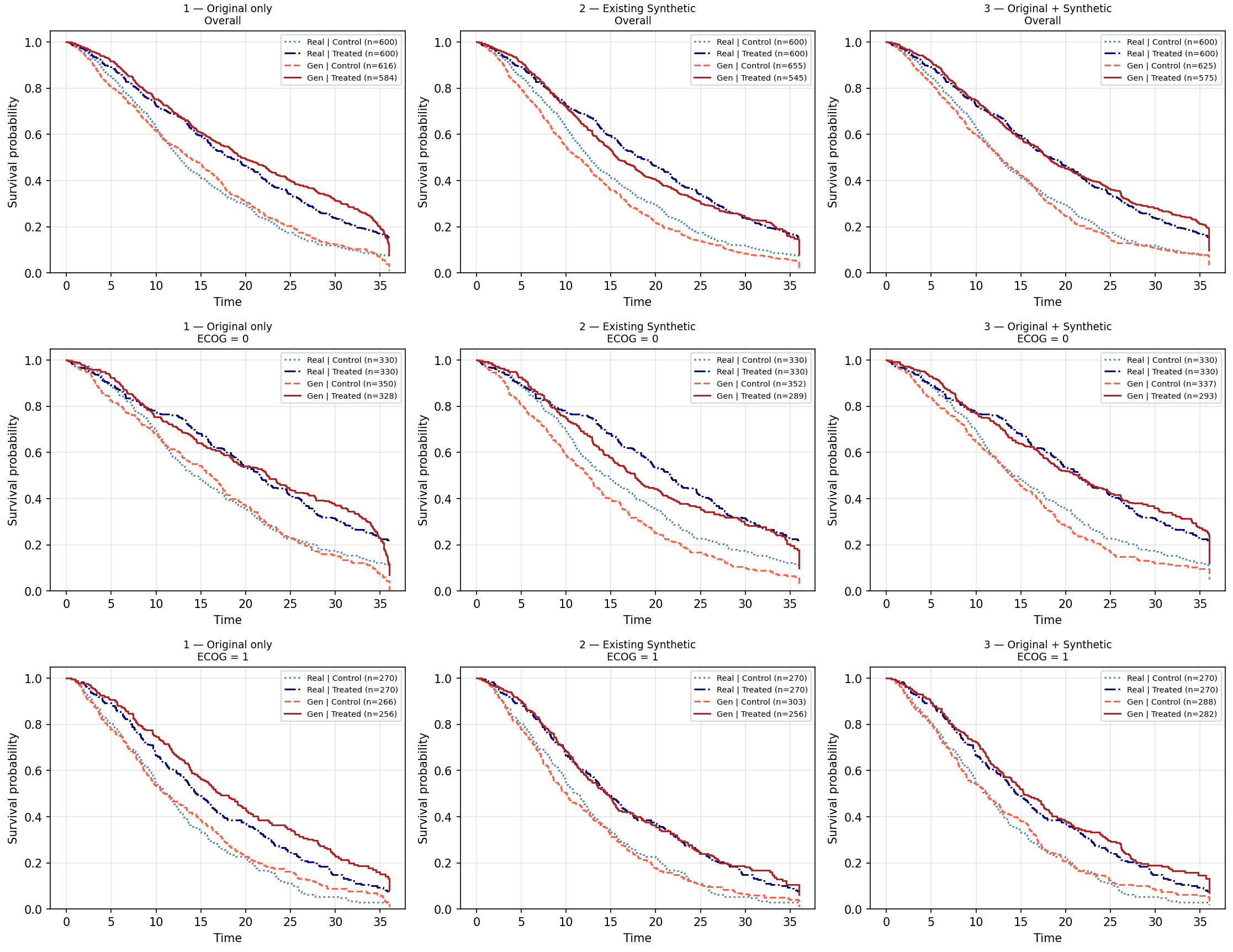} \caption{KM diagnostic plots comparing the real IPD and SurvDiff-generated datasets under three training scenarios: original IPD, SynthIPD, and IPD + SynthIPD combined. ECOG is the hypothetical subgroup name taking values in $\{0,1\}$.} \label{fig:km_survdiff} \end{figure}

\begin{figure}[p] \centering \includegraphics[width=\textwidth]{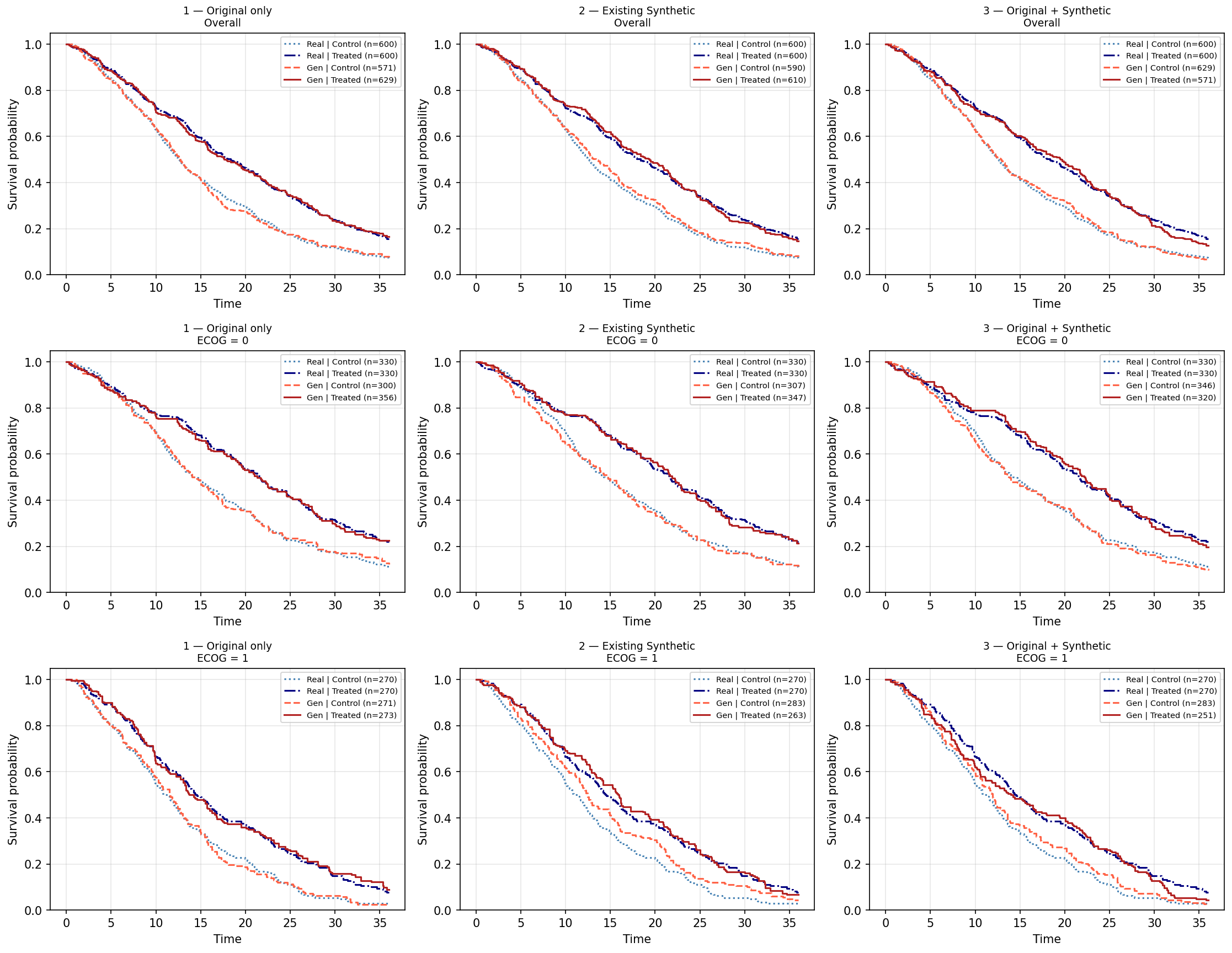} \caption{KM diagnostic plots comparing the real IPD and conditional tree generated datasets under three training scenarios: original IPD, SynthIPD, and IPD + SynthIPD combined. ECOG is the hypothetical subgroup name taking values in $\{0,1\}$.} \label{fig:km_tree} \end{figure}

In addition to the C-index discrepancy analysis in Table~\ref{tab:cindex_discrepancy_models}, we provide KM diagnostic plots in Figures~\ref{fig:km_survivalgan}-\ref{fig:km_tree}, for each downstream generative model. For each model, the generated datasets under the three training scenarios are compared with the original IPD, both overall and within-strata. 

\section{Additional information for Subsection \ref{subsec:n9741}}\label{app:additional info}

\begin{figure}[htbp]
    \centering
    \includegraphics[width=\linewidth,height=0.5\textheight]{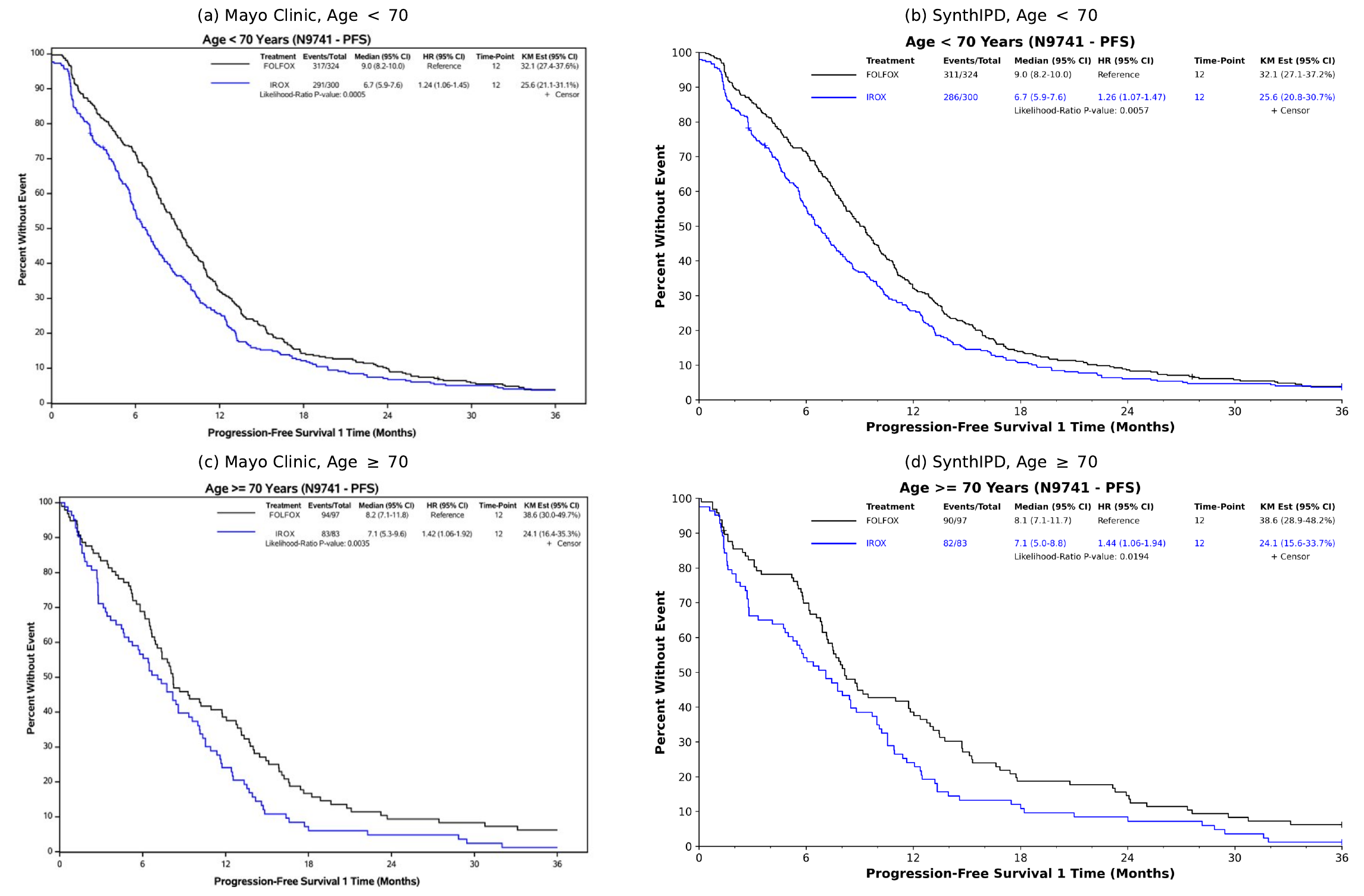}
  %     % Row 3
  % \begin{minipage}[t]{0.48\textwidth}
  %   \centering
  %   \includegraphics[width=\linewidth,height=0.18\textheight]{n9741_ageless70_mayo.png}
  %   \textbf{(a)}
  % \end{minipage}\hfill
  % \begin{minipage}[t]{0.48\textwidth}
  %   \centering
  %   \includegraphics[width=\linewidth,height=0.18\textheight]{n9741_ageless70_synthetic.png}
  %   \textbf{(b)}
  % \end{minipage}

  % \vspace{0.3em}

  % % Row 4
  % \begin{minipage}[t]{0.48\textwidth}
  %   \centering
  %   \includegraphics[width=\linewidth,height=0.18\textheight]{n9741_agegreater70_mayo.png}
  %   \textbf{(c)}
  % \end{minipage}\hfill
  % \begin{minipage}[t]{0.48\textwidth}
  %   \centering
  %   \includegraphics[width=\linewidth,height=0.18\textheight]{n9741_agegreater70_synthetic.png}
  %   \textbf{(d)}
  % \end{minipage}
    \caption{The subgroup KM plots (a)-(d) stratified by age$\geq 70$ or age$<70$. The left $2$ plots are provided by Mayo Clinic for comparison purpose only. The right $2$ plots are generated by SynthIPD. Two plots in the same row correspond to the same subgroup.}
    \label{fig:n9741 age}
\end{figure}

In this section, we complement results not provided in Section \ref{sec:case studies}. Recall in Subsection \ref{subsec:n9741}, the survival information for the gender subgroup is reconstructed. Here, we consider another subgroup, age. The SynthIPD KM plots are almost identical to the true KM plots (Figure~\ref{fig:n9741 age}). The mRAEs have magnitude less than $1\%$ for both median and HR. 

\end{document}